\documentclass[prodmode,acmtecs]{acmsmall} 

\usepackage[ruled]{algorithm2e}

\SetAlFnt{\small}
\SetAlCapFnt{\small}
\SetAlCapNameFnt{\small}
\SetAlCapHSkip{0pt}
\IncMargin{-\parindent}

\acmVolume{9}
\acmNumber{4}
\acmArticle{39}
\acmYear{2014}
\acmMonth{1}
\usepackage{color}

\usepackage{subfig}
\usepackage{graphicx}
\usepackage{caption}

\begin{document}

\markboth{C. A. Schwaninger et al.}{Displacement Field Algorithms for Morphogenesis}

\title{
 Simulating Organogenesis: Algorithms for the Image-based Determination of Displacement Fields}
\author{CLEMENS ARTHUR SCHWANINGER
\affil{ETH Zurich}
DENIS MENSHYKAU
\affil{ETH Zurich}
DAGMAR IBER
\affil{ETH Zurich}}

\begin{abstract}
Recent advances in imaging technology now provide us with 3D images of developing organs. These can be used to extract 3D geometries for simulations of organ development. To solve models on growing domains, the displacement fields between consecutive image frames need to be determined. Here we develop and evaluate different landmark-free algorithms for the determination of such displacement fields from image data. In particular, we examine minimal distance, normal distance, diffusion-based and uniform mapping algorithms and test these algorithms with both synthetic and real data in 2D and 3D. We conclude that in most cases the normal distance algorithm is the method of choice and wherever it fails, diffusion-based mapping provides a good alternative. 
\end{abstract}

\acmformat{C. Arthur Schwaninger, Denis Menshykau, Dagmar Iber, 2014. Simulating Organogenesis: Algorithms for the Image-based Determination of Displacement Fields}


\begin{bottomstuff}
This work is supported by SystemsX grants of the Swiss National Fund (SNF).

Author's addresses: C.A. Schwaninger, D. Menshykau  {and} D. Iber,
D-BSSE, ETH Zurich, Mattenstrasse 26, 4058 Basel, Switzerland; Swiss Institute of Bioinformatics (SIB), Switzerland.
\end{bottomstuff}

\maketitle

\section{Introduction}

During the development of multicellular organisms, patterns emerge that result in the  appearance of body axes, organs, appendages and other functional structures. While many of the responsible genes have been identified, it is still unclear how the patterning processes are controlled in time and space and the regulatory processes are typically too complex to be addressed by verbal reasoning alone. Mathematical models have a long history in developmental biology \cite{Turing:1952,Wolpert:1969p21589}, but have long been difficult to test experimentally.  Recent technical advances now yield an increasing amount of quantitative data \cite{Oates:2009p28318,Wartlick:2009p49377,Wartlick:2011p49380}, and now permit the generation of more detailed, data-based and testable models \cite{Iber:2012hm,Morelli2012}. \\
 
We and others have developed mathematical models for a range of developmental processes as recently reviewed in \cite{Iber:OpenBiol:2013,Iber:BirthDefectsResCEmbryoToday:2014}, including the control of branching morphogenesis \cite{Menshykau:2012kg,Celliere:2012jc,Menshykau:nMxfL07C}, limb development \cite{Probst:2011jo,Badugu:2012ho} and bone development \cite{Tanaka:2013wt}.  The models are based on partial differential equations (PDEs) of the form
\begin{equation}{\label{eq:RD}}
\dot{c}_\mathrm{i} =D_\mathrm{i}\Delta c_\mathrm{i}+R(c_1, \, \ldots, c_n)
\end{equation}
where $\dot{c}_\mathrm{i}$ denotes the time derivative of $c_i$, $c_i$ denotes the concentration of the different species $i$, $D_i$ represents the diffusion coefficient of species $i$,  $\Delta$ the  Laplace operator, and $R(c_1, \, \ldots, c_n)$ the biochemical reactions between the $n$ different species $i \in [1, n]$. Parameter values for the models are set so that embryonic gene expression and signalling patterns from wild type and mutant mice can be reproduced on the idealised 2D and 3D domains. To further increase the accuracy and thus the predictive power of the models, it is important to solve them on realistic embryonic geometries. Recent advances in imaging technology, algorithms and computer power now permit the development of such increasingly realistic simulations of biological processes. In particular, it is now possible to obtain 2D and 3D shapes of developing organs and to solve the models on those embryonic geometries \cite{Gleghorn:2012el,Clement:2012fj,Menshykau:2014,Iber:2015dz}.\\

\begin{figure}[t]   
 \centering
 \includegraphics[width=0.98\textwidth]{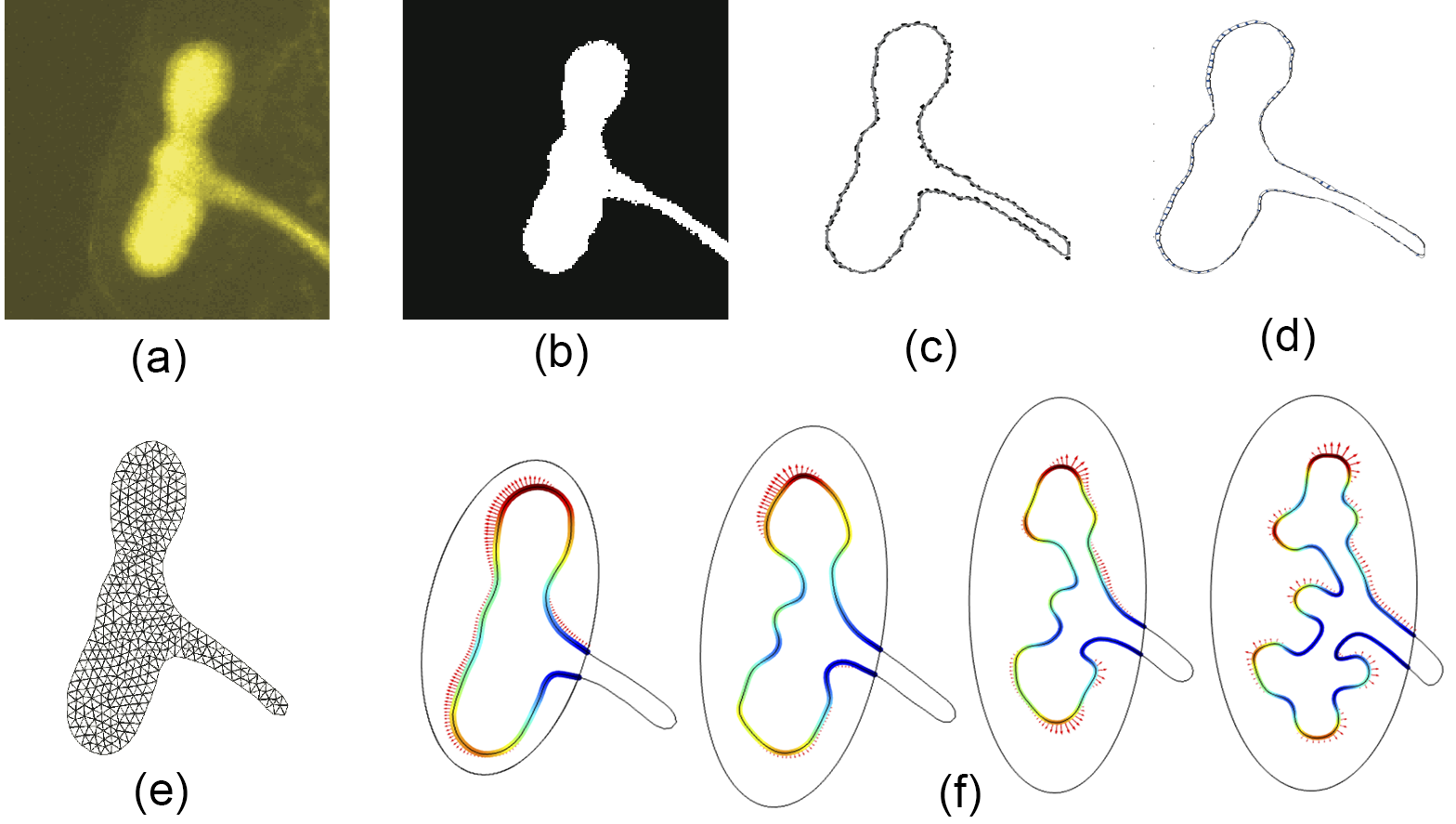}%
 \caption{\textbf{The Image-Based Modelling Approach.} a) an image of mouse embryonic kidney explant, b) segmented image, c) extracted border in the shape of the embryonic kidney explant, d) displacement field, e) computational mesh and f) solution of a computational model on a growing domain. Adapted from \protect\cite{Adivarahan:2013iz}.\label{fig:intro}}
\end{figure}

During embryonic development, growth and patterning are tightly linked \cite{Iber:2015dz}. It is therefore insufficient to solve the reaction-diffusion equations on a static domain. When solving Eq.\,\ref{eq:RD} on a growing domain, we need to include advection and dilution terms \cite{Iber:2015dz} and obtain 
\begin{equation}{\label{eq:RD_growth}}
\dot{c}_\mathrm{i}+\nabla(\mathbf{u} \cdot c_\mathrm{i})=D_\mathrm{i}\Delta c_\mathrm{i}+R(c_1, \, \ldots, c_n).
\end{equation}
Here, $\mathbf{u}$ represents the velocity field of the moving domain. To develop a mechanistic model of growth control that yields $\mathbf{u}$, we would need to understand how growth is controlled at the molecular level during development. This is not yet the case, and recent measurements demonstrate  that simple growth models \cite{Dillon:JTheorBiol:1999} fail to correctly predict shape changes during developmental growth \cite{Boehm:PlosBiol:2010}. Alternatively, one could measure $\mathbf{u}$ for biological systems by tracking cell division and cell movements in the growing domain. So far, this has been done only in very few experimental systems, e.g. \cite{Bellaiche2011}. To obtain an estimate of $\mathbf{u}$ for our experimental systems of interest, we therefore use imaging data at different stages of the deformation process to estimate the local rates of tissue deformation \cite{Adivarahan:2013iz}. An example of a snapshot of such a 2D time series is shown in Figure\,\ref{fig:intro}a.  To extract the local rates of tissue deformation from the available imaging data, we segment the images (Fig. \ref{fig:intro}b) and extract the borders in 2D or surfaces in 3D from the images (Fig. \ref{fig:intro}c). We then need to determine a displacement field that maps the earlier border or surface on the later border or surface (Fig.\,\ref{fig:intro}d). Once we have this displacement field between the boundaries of the two subsequent geometries, we can use this in our simulations. To this end, we read the initial geometry and the displacement field into the PDE solver (in our case COMSOL \cite{Menshykau_COMSOL2012}). We then mesh the computational domain (Fig.\,\ref{fig:intro}e). Finally, we solve the model given by Eq.\,\ref{eq:RD_growth} on the embryonic, growing domain (Fig.\,\ref{fig:intro}f). While solving the model, the PDE solver deforms the boundary according to the displacement field, and passively stretches the domain to follow the deforming boundary. The deformation inside the domain yields the velocity field $\mathbf{u}$. It is important to note that in this way, we obtain a deformation that recapitulates the embryonic shape changes. However, since experimental data is missing that would yield information on the movement of boundary and internal points, both the mapping of the boundary points and the deformations inside the domain, and thus the velocity field $\mathbf{u}$, are arbitrary and therefore do not necessarily correspond to the physiological changes. For sufficiently small deformations, i.e. for sufficiently small time steps between boundaries, the error can, however, be expected to be sufficiently small such that we still obtain useful results.\\

Since the choice of displacement field is arbitrary, it is desirable that the mapping algorithm delivers a displacement field that facilitates simulations and that works automatically without any further user input. Many algorithms have been proposed  to address this issue, e.g. \cite{Lazarus:VisualComputer:1998,Alexa2002,Sigal2008,Athanasiadis2010}. However, algorithms that work also for non-convex surfaces  (such as branching geometries) without user input are rare, and will be computationally costly if generally applicable \cite{Lazarus:VisualComputer:1998}. The commercial software package AMIRA, which we used previously \cite{Menshykau:2014}, offers the landmark based Bookstein algorithm \cite{Bookstein1989}. However, the Bookstein algorithm requires the manual identification of points of correspondence, so-called landmarks,  which leads to \textit{ad hoc} rather than algorithmic displacement fields and requires a lot of manual work. We have previously also used a  minimal distance based algorithm, which is straight-forward to implement, but which frequently fails to deliver useful displacement fields and then requires manual curation \cite{Adivarahan:2013iz}.  To overcome these drawbacks, we developed
four basic algorithms and their extensions that can be used to calculate the displacement field between two consecutive stages without any user input. We introduce a quality measure to evaluate our algorithms and test the algorithms with image data for kidney and lung branching morphogenesis.

 \begin{figure}[t]   
  \centering
  \subfloat[Displacement field of entire domain]{\label{fig:kidney_normal_intro}%
  \includegraphics[width=0.4\textwidth]{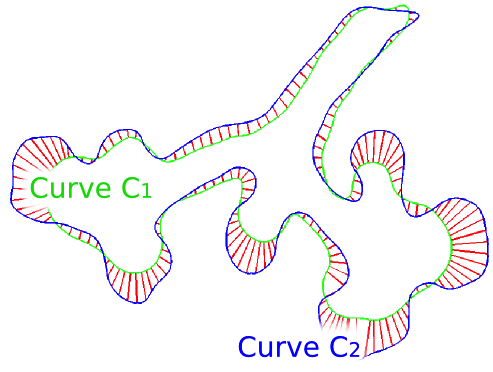}}%
  \quad%
 \subfloat[Local displacement field]{\label{fig:displacement_field}%
  \includegraphics[width=0.4\textwidth]{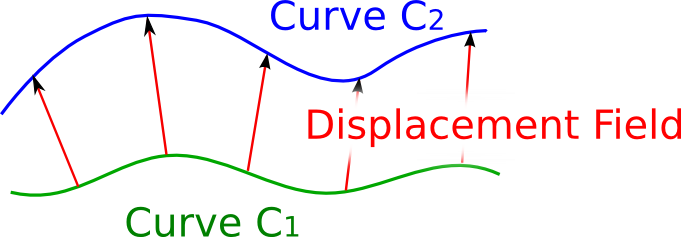}}
  \quad%
  \caption{ \textbf{Displacement Field.} Curve $C_1$ represents the shape of an embryonic structure at a time step $t$; curve $C_2$ represents the same structure at a later time point $t + \Delta t$. For our simulations of growing organs it is essential to find a high quality displacement field that maps  points from $C_1$ to $C_2$. (a) An example of the displacement field for the entire growing embryonic kidney explant. (b) The displacement field at some local position of the domain shown in panel (a).}
  \label{fig:intro_kidney}
\end{figure}

\section{Algorithms for Computing the Displacement Fields}

\subsection{Basic Algorithms} 
 \label{sec:basic_algs}

We denote the curve at time $t$ by $C_1$ and the curve at time $t + \Delta t$ by $C_2$ as shown in Figure\,\ref{fig:intro_kidney}.  Prior to the calculation of the displacement field we interpolated the set of discrete points on $C_1$ and $C_2$ such that both curves contain the same number of points that were equally spaced on each curve. Here we used the MATLAB function {\tt interparc} \cite{interparc} and we used spline interpolation for smooth curves and linear interpolation in case of corner points. The interpolated points were then mapped using one of the following basic algorithms. In Section \ref{sec:extensions} 
we will show how these algorithms can be extended.

\begin{enumerate}
  \item \label{alg:mindist} \textit{Minimal Distance Mapping:} Every point on the curve $C_1$ is mapped to the point on the curve $C_2$ to which it has the minimal distance. The point on $C_2$ can be determined using the MATLAB function {\tt distance2curve} \cite{distance2curve}. 
 
  \item \label{alg:normal} \textit{Normal Mapping:} The normal vector is determined at every
  point on $C_1$ and is mapped to the closest point on $C_2$ where the normal and $C_2$ intersect.
  \item \label{alg:diff} \textit{Diffusion Mapping:} This mapping is obtained by solving the steady state diffusion equation (Laplace equation)
 \begin{equation}
 \Delta c = 0,  {\label{eq:Laplace}}
\end{equation}
which follows from Eq. \ref{eq:RD} in steady state ($\dot{c} = 0$) and in the absence of reactions ($R = 0$). Eq.\,\ref{eq:Laplace} is solved in the area between the two curves in 2D, or between two surfaces in 3D using COMSOL Multiphysics (Fig.\,\ref{fig:diff_solution}). As initial conditions we use Dirichlet boundary conditions and we set the value $c$ to be $1$ on $C_1$ and $0$ on $C_2$ or \textit{vice versa}. The start and end points on $C_1$ and $C_2$ are determined by computing the streamlines using particle tracking (Fig.\,\ref{fig:diff_streamlines}); the start and end points define the mapping vector (Fig.\,\ref{fig:diff_mapping_example}). 
\begin{figure}[t]   
\begin{center}
 \subfloat[Solution of the diffusion equation]{\label{fig:diff_solution}%
    \includegraphics[trim=1cm 1.2cm 1cm 1.2cm, clip=true, width=0.33\textwidth]{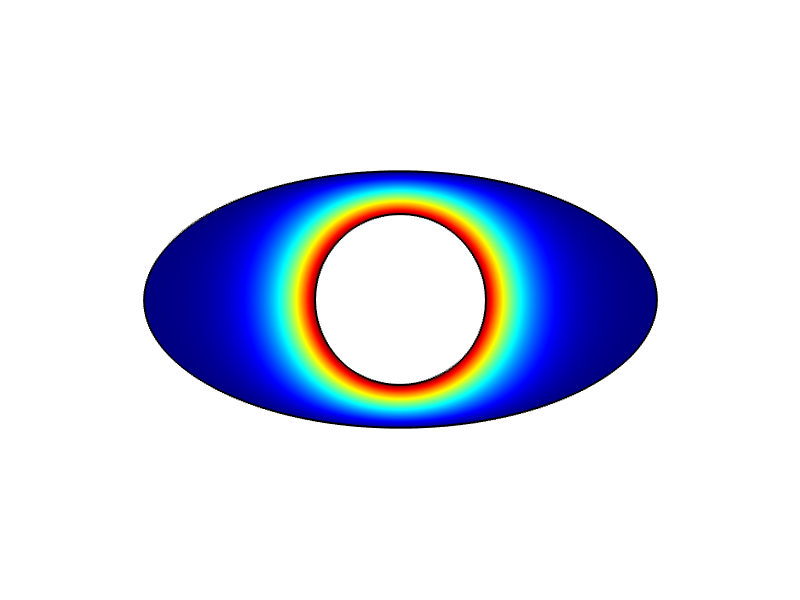}}
   \subfloat[Streamlines]{\label{fig:diff_streamlines}%
    \includegraphics[trim=1cm 1.2cm 1cm 1.2cm, clip=true, width=0.33\textwidth]{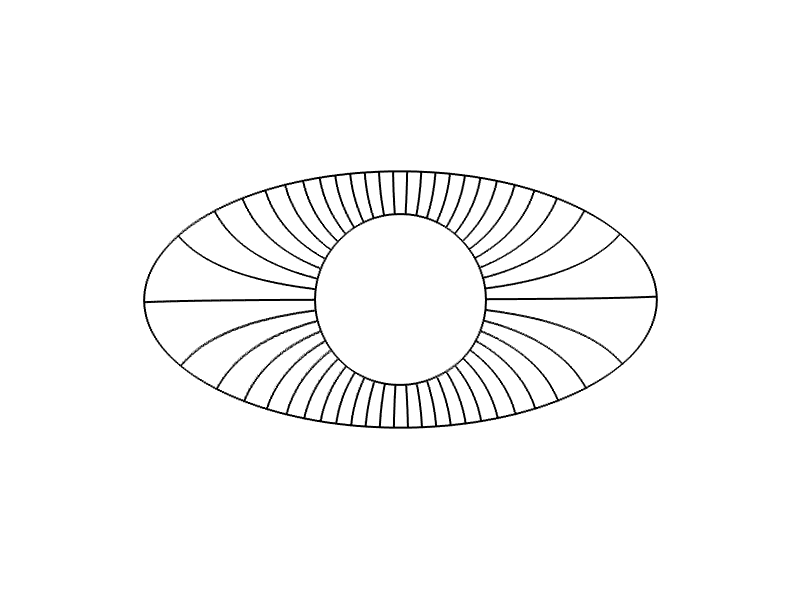}}%
   \subfloat[Diffusion mapping]{\label{fig:diff_mapping_example}%
    \includegraphics[trim=1.6cm 3.2cm 0cm 1cm, clip=true, width=0.37\textwidth]{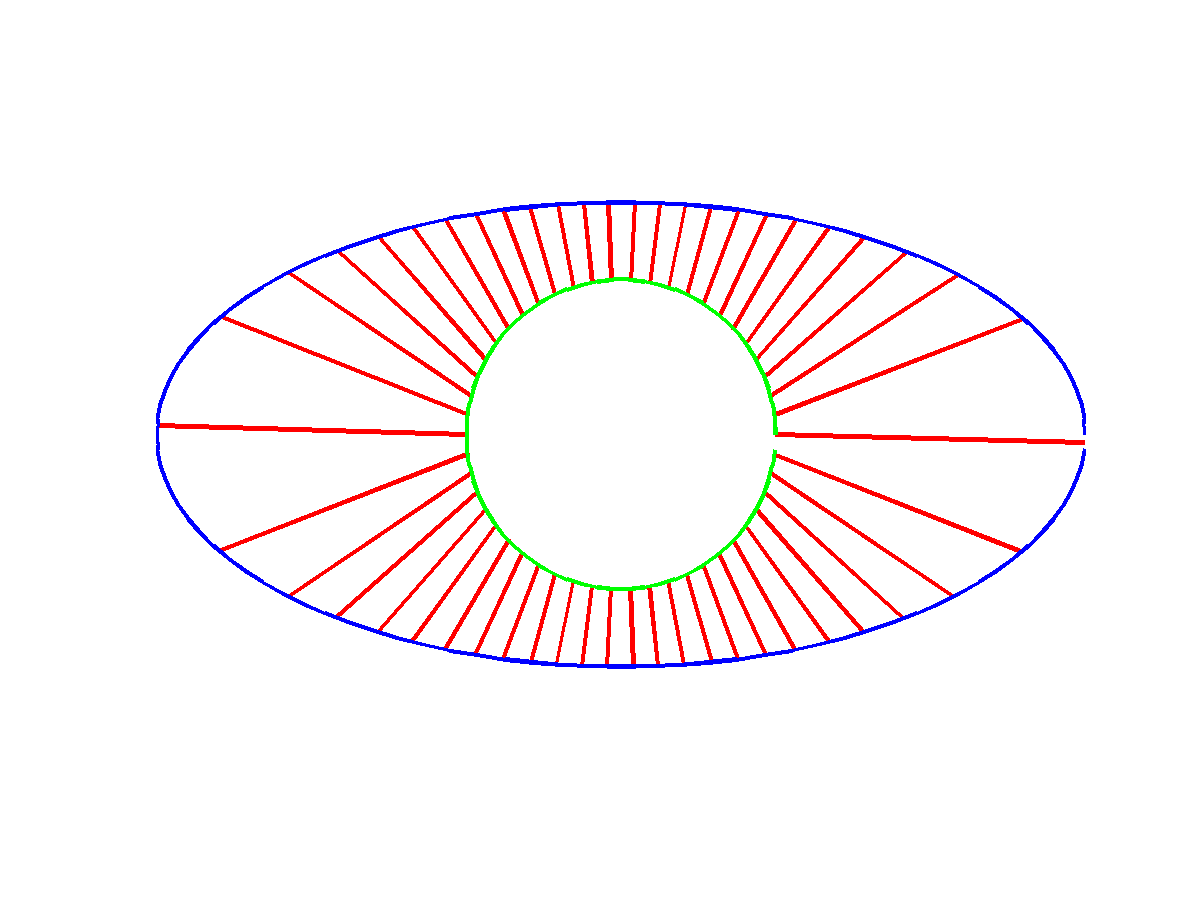}}%
 \end{center}
 \caption{\textbf{Diffusion Mapping}. (a) The numerical solution of the diffusion equation. (b) The streamlines, as obtained by particle tracking from $C_1$ to $C_2$. (c) The displacement field vectors are obtained by connecting start and end points of the streamlines. \label{fig:diffusion_explained}}
\end{figure}

\item \textit{Uniform Mapping:} For this algorithm we need at least one point on $C_1$ and one point on $C_2$, for  which we know that they should be mapped onto each other. These two points can either be selected manually or algorithmically. To determine such points algorithmically, one could use curve intersection points in case of intersecting curves, or in case of open, non-intersecting curves, beginning and end points. If the mapping  has been defined for at least one single point, we can interpolate $N$ points equidistantly along both curves, starting from the predefined point and ending at the same point (for closed curves) or at another predefined point (for open curves or curve segments). We used the MATLAB function {\tt interparc} to calculate the $N$ equidistant points on a curve.
\end{enumerate}

\subsection{Extensions} 
\label{sec:extensions}
In some cases, prior to computing the displacement field with one of the basic algorithms, the following additional steps can improve the quality of the resulting displacement field.

\begin{figure}[t]   
\begin{center}
  \subfloat[Global mapping]{\label{fig:curve_segments1}%
    \includegraphics[width=0.45\textwidth]{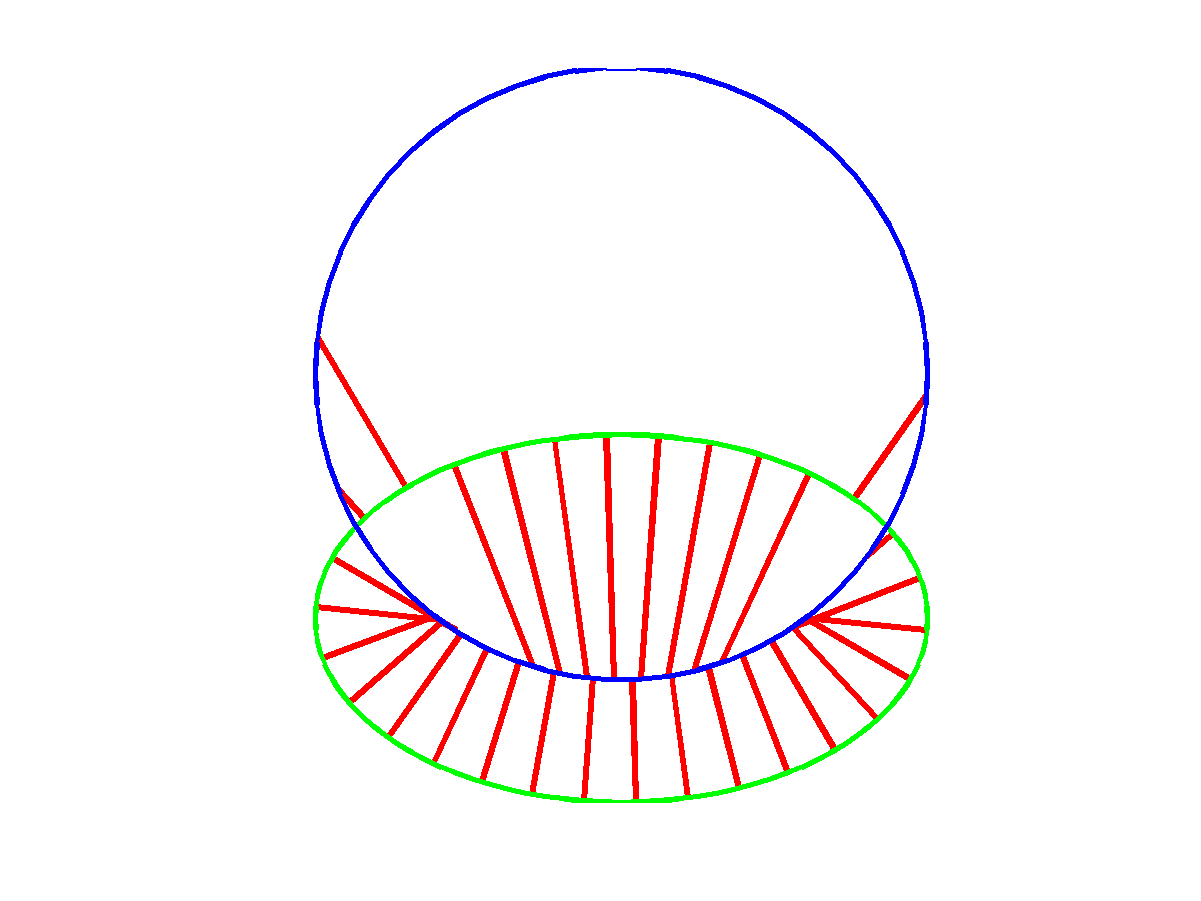}}%
 \quad%
 \subfloat[Segment mapping]{\label{fig:curve_segments2}%
    \includegraphics[width=0.45\textwidth]{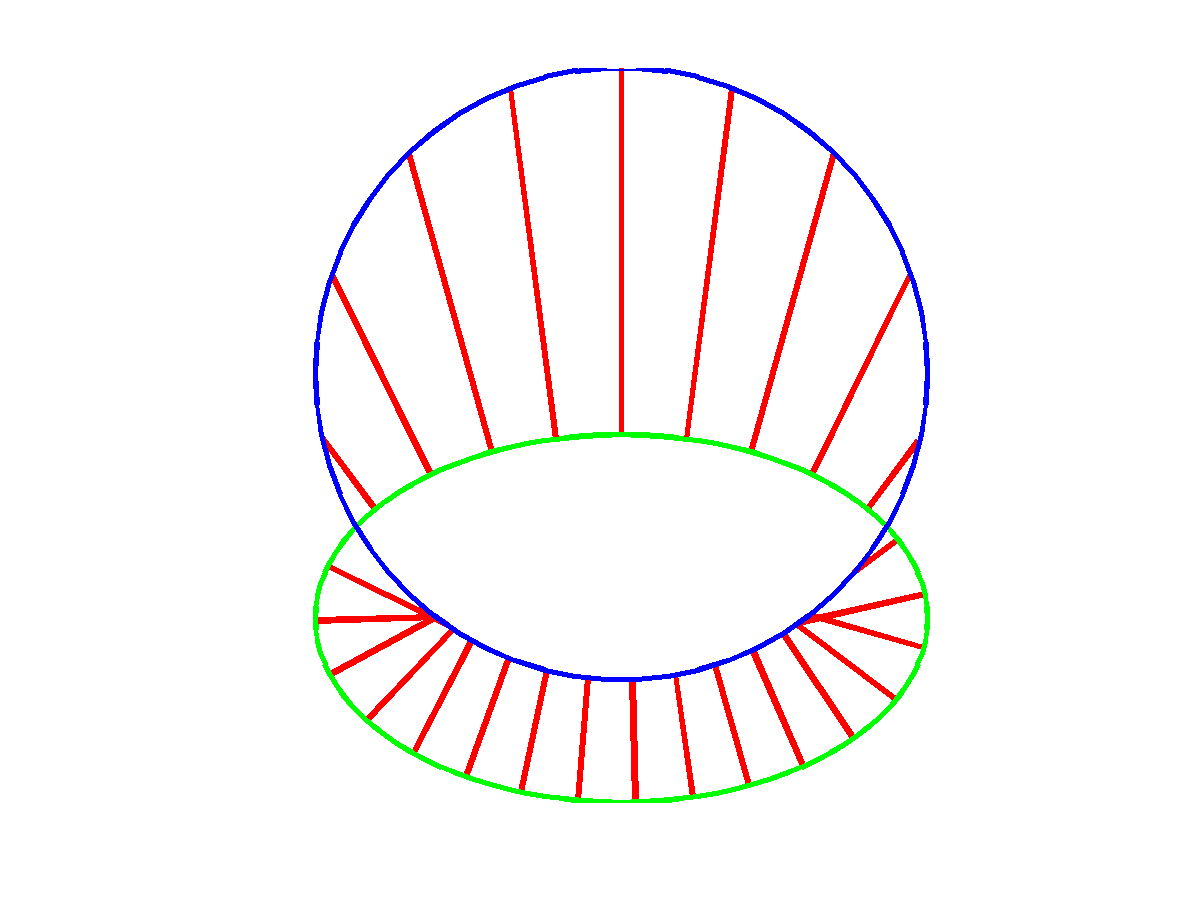}}
 \quad%
 \caption{\textbf{Mapping of Curves with Intersections}. (a) The displacement field was computed globally using normal mapping. Note that many points map to the wrong segment on $C_2$. (b) The curves were split into subdomains based on the intersection points and only segments were mapped onto each other using normal mapping. The mappings are now correct. \label{fig:curve_segments}}
 \end{center}
\end{figure}

\begin{enumerate}
  \item \textit{Reverse Mapping:} 
  The mapping can be done from curve $C_2$ to $C_1$.
  \item \textit{Transformation of $C_1$:} \label{sec:transformation}
  \label{transform} 
The more similar the two curves $C_1$ and $C_2$ are, the easier it is to find a good displacement field and the less it depends on which of the four above mentioned algorithms we choose. Examples include cases where not only the shape, but also the lengths of the curves differ. It can then be beneficial to first perform linear operations that transform the curve $C_1$ to curve $C_{1,\mathrm{t}}$, from which we then compute the displacement field $\mathcal{D}_\mathrm{t}$. Finally, we invert the transformation and compute the displacement field $\mathcal{D}$ for the original curves $C_1$ to $C_2$. A detailed description of our algorithm can be found in Appendix A.

  \item \textit{Curve Segment Mapping} \label{segment_mapping} Curves $C_1$ and $C_2$ can have intersection points (Fig.\,\ref{fig:kidney_normal_intro}) and it is then often beneficial to first split the domain based on the intersection points into a set of subdomains, where the basic mapping routines are carried out independently.  Figure \ref{fig:curve_segments} shows the displacement fields that were computed with the normal mapping algorithm when intersection points were either ignored (Fig.\,\ref{fig:curve_segments1}) or used to split the domain into independent subdomains (Fig.\,\ref{fig:curve_segments2}). If the intersection points are ignored, then many points are mapped in the wrong direction because the algorithm maps to the closest normal intersection point. This issue can be resolved if curves are split into segments according to the intersection points (Fig.\,\ref{fig:curve_segments2}).
\end{enumerate}

\section{Evaluation}

\subsection{Qualitative Evaluation of the Mapping Algorithms} 

\begin{figure}[h]   
\begin{center}
  \subfloat[Minimal distance mapping]{\label{fig:circle_min}%
    \includegraphics[trim=2cm 3cm 1.5cm 1cm, clip=true, width=0.33\textwidth]{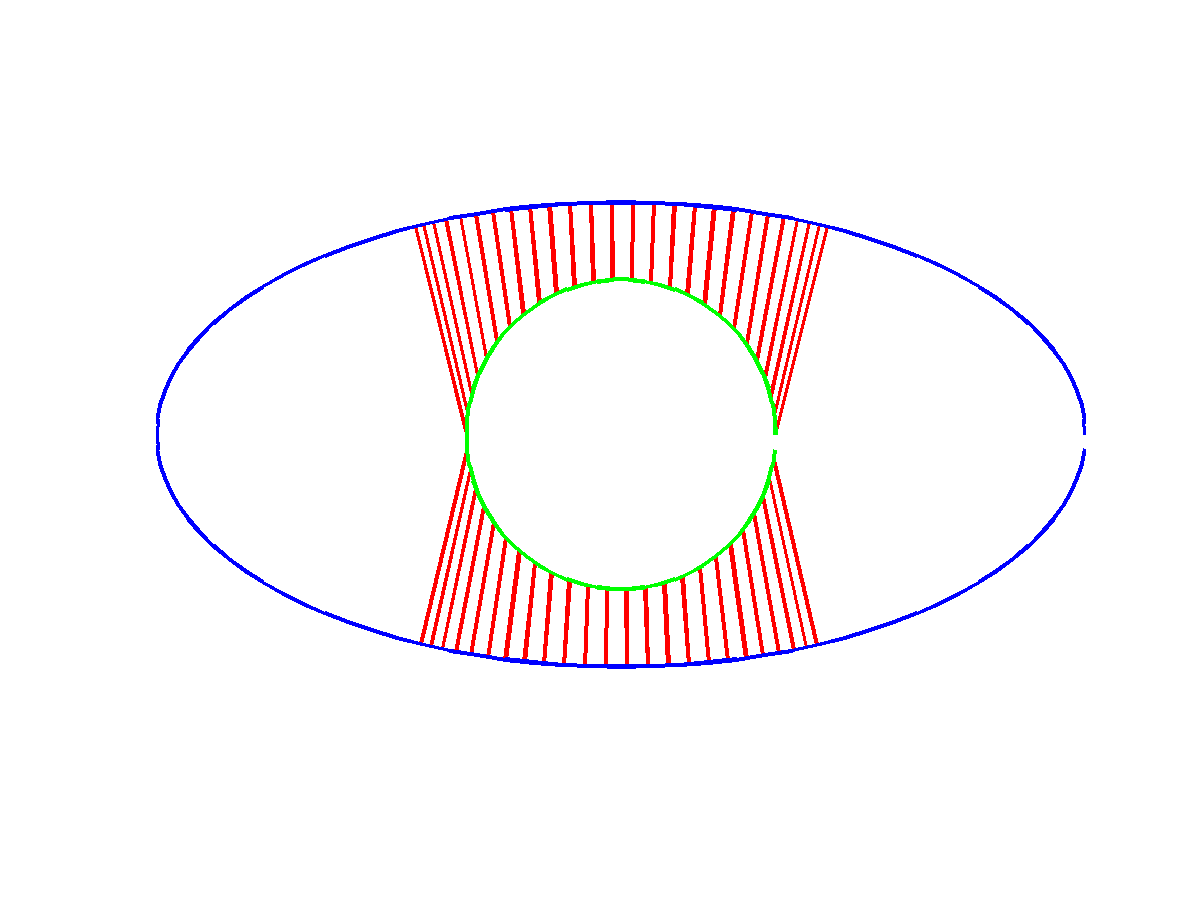}}%
   \subfloat[Reverse minimal distance mapping]{\label{fig:circle_rev_min}%
    \includegraphics[trim=2cm 3cm 1.5cm 1cm, clip=true, width=0.33\textwidth]{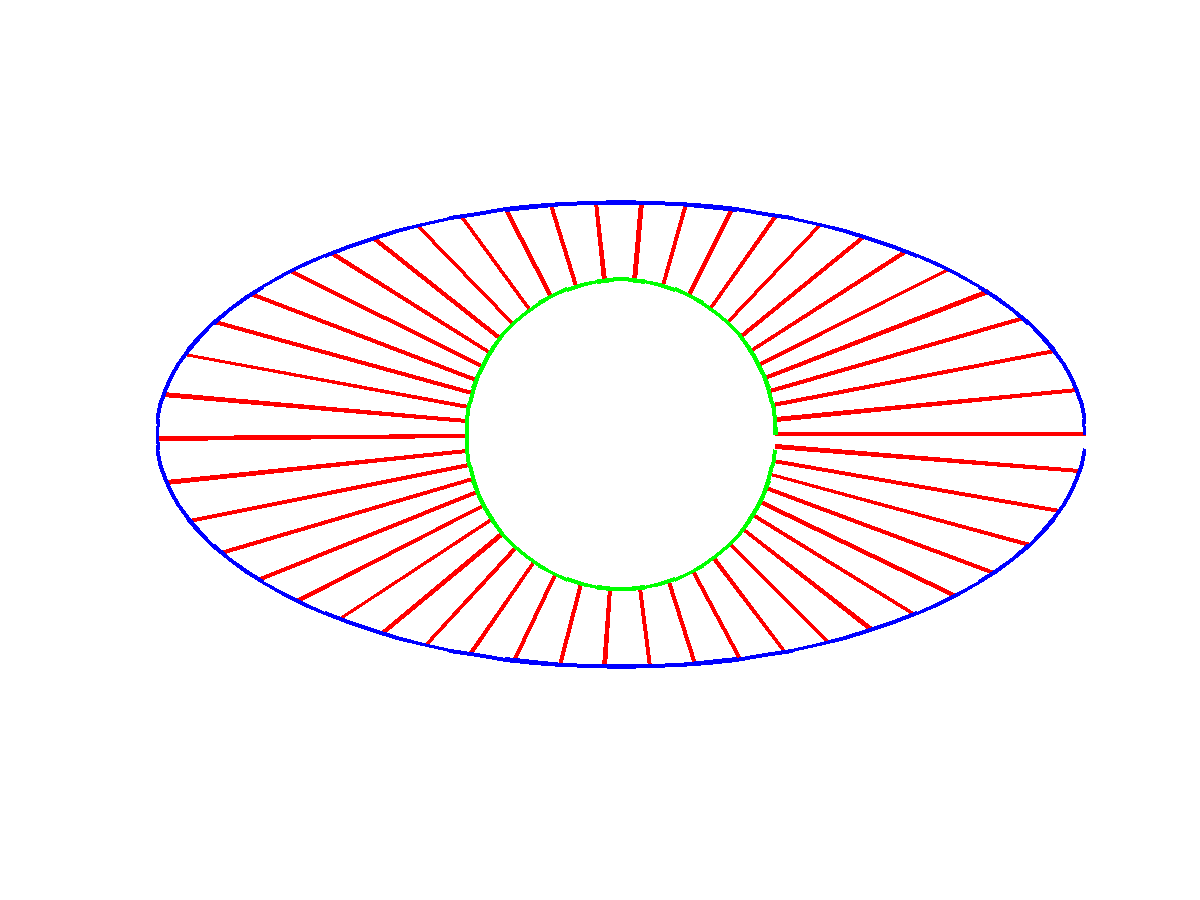}}%
   \subfloat[Transformed minimal distance mapping]{\label{fig:circle_trans_min}%
    \includegraphics[trim=2cm 3cm 1.5cm 1cm, clip=true, width=0.33\textwidth]{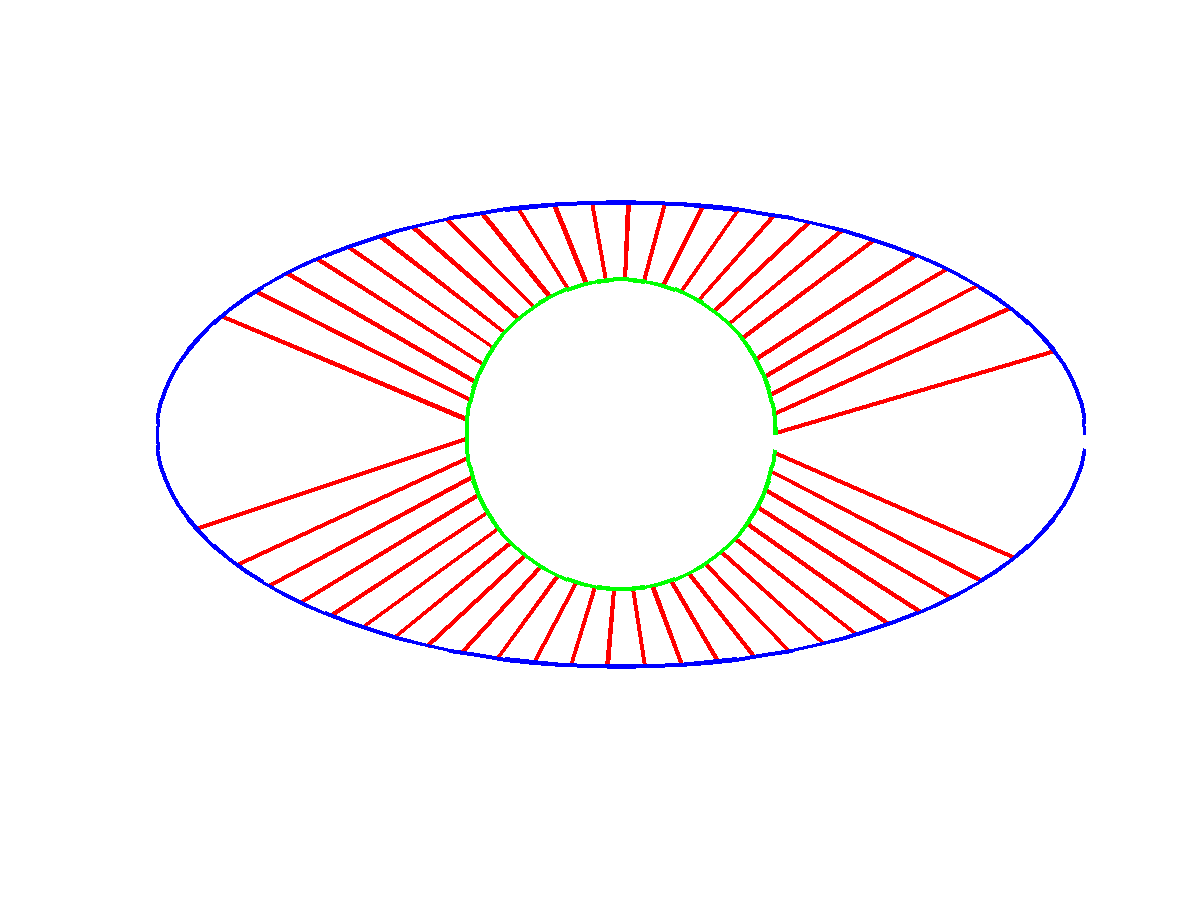}}%
 \quad%
 \subfloat[Normal mapping]{\label{fig:circle_normal}%
    \includegraphics[trim=2cm 3cm 1.5cm 1cm, clip=true, width=0.33\textwidth]{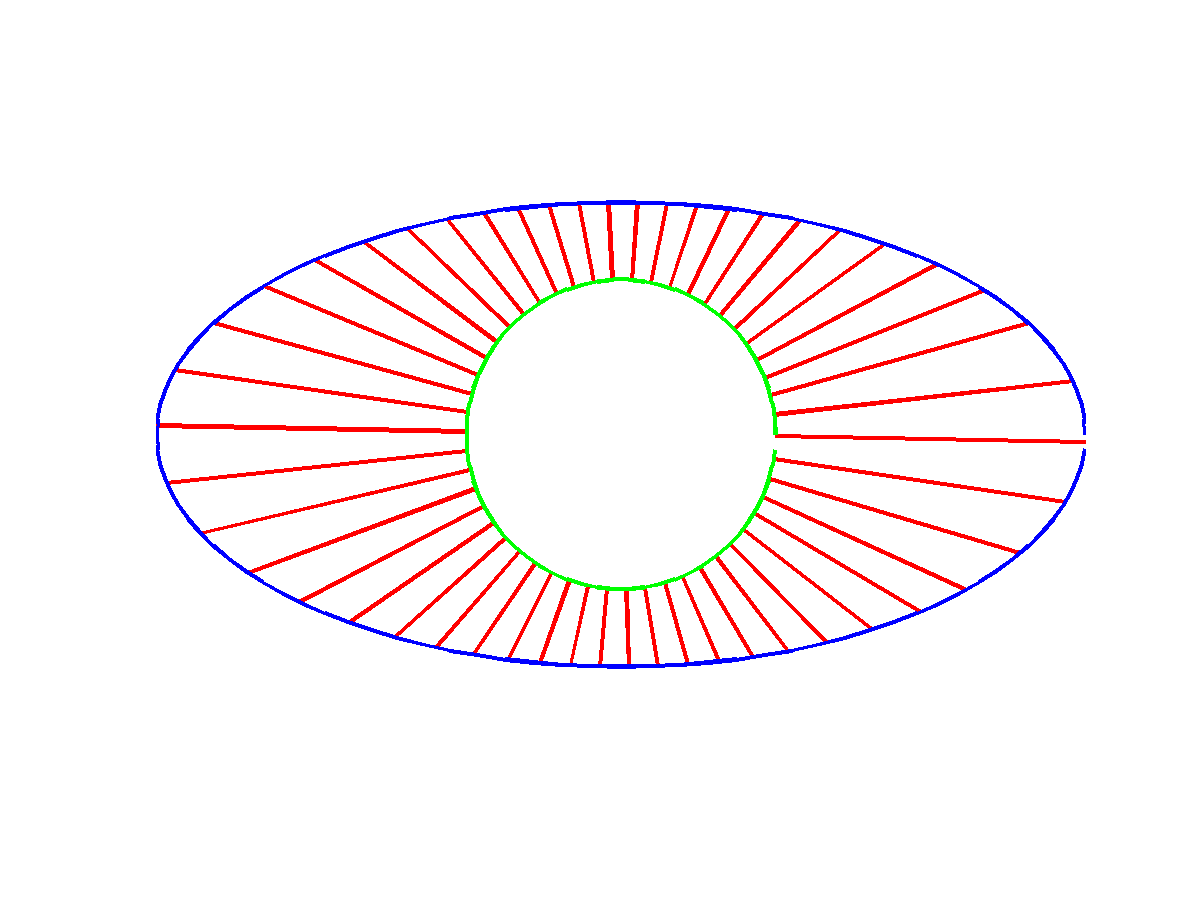}}
  \subfloat[Reverse normal mapping]{\label{fig:circle_rev_normal}%
    \includegraphics[trim=2cm 3cm 1.5cm 1cm, clip=true, width=0.33\textwidth]{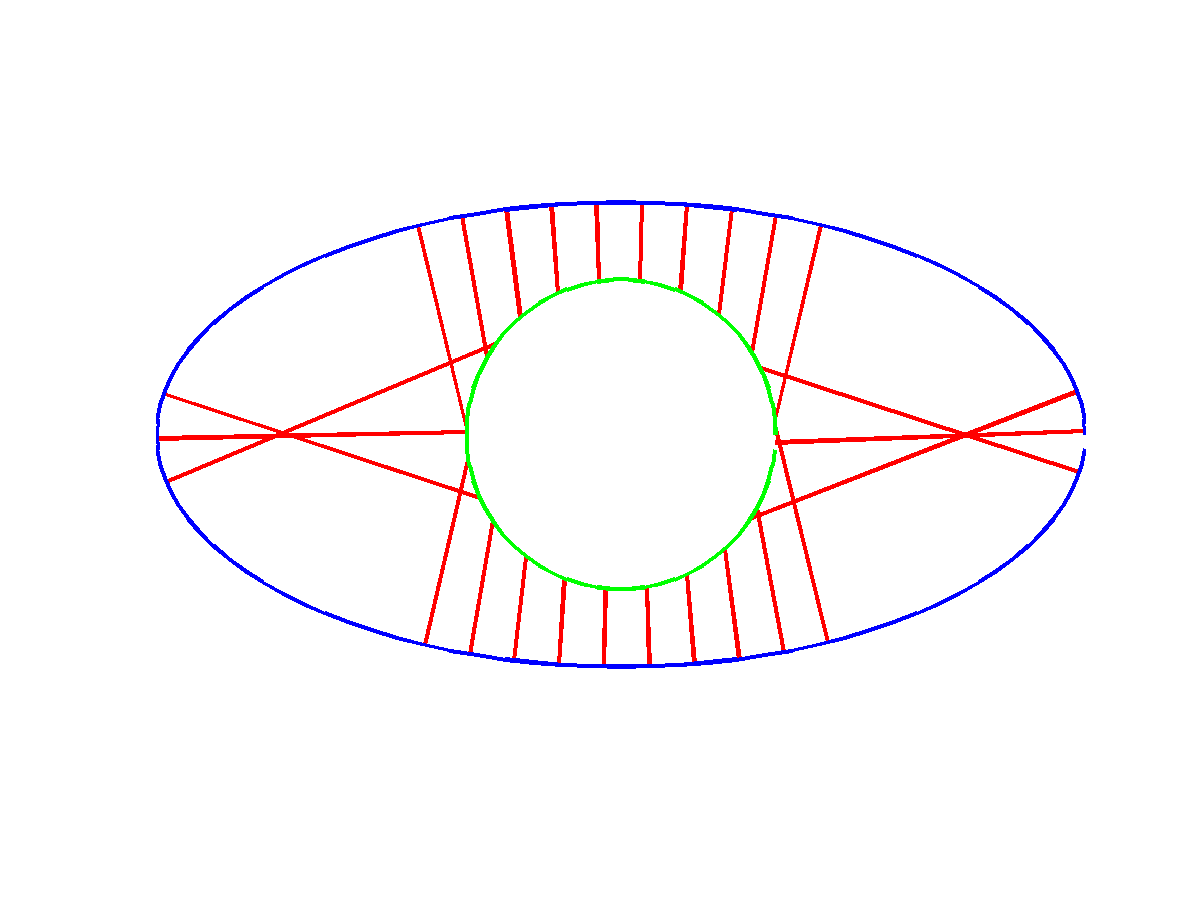}}
  \subfloat[Transfomed normal mapping]{\label{fig:circle_trans_normal}%
    \includegraphics[trim=2cm 3cm 1.5cm 1cm, clip=true, width=0.33\textwidth]{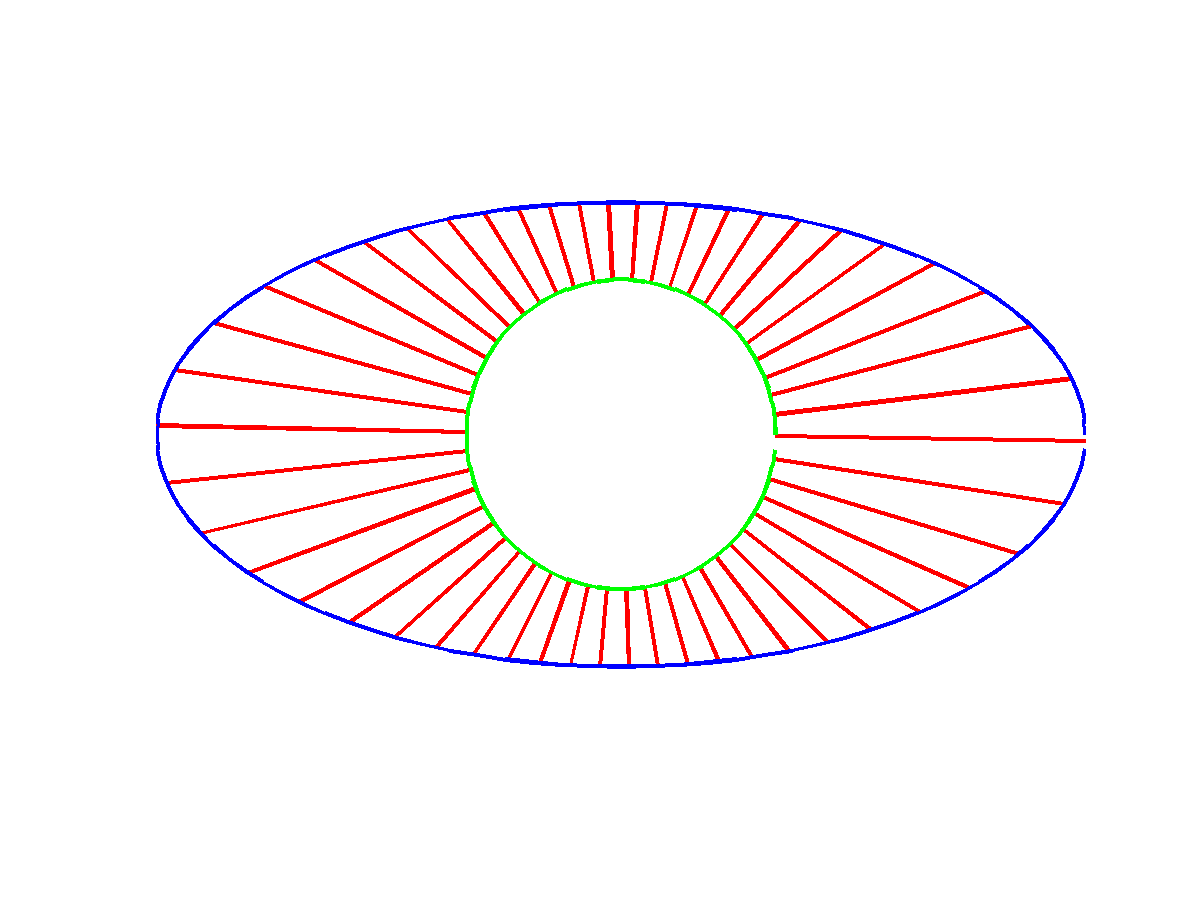}}
 \quad%
 \subfloat[Diffusion mapping]{\label{fig:circle_diff}%
    \includegraphics[trim=2cm 3cm 1.5cm 1cm, clip=true, width=0.33\textwidth]{images/circle_diffusion_mapping_100points.png}}
   \subfloat[Reverse diffusion mapping]{\label{fig:circle_rev_diff}%
    \includegraphics[trim=2cm 3cm 1.5cm 1cm, clip=true, width=0.33\textwidth]{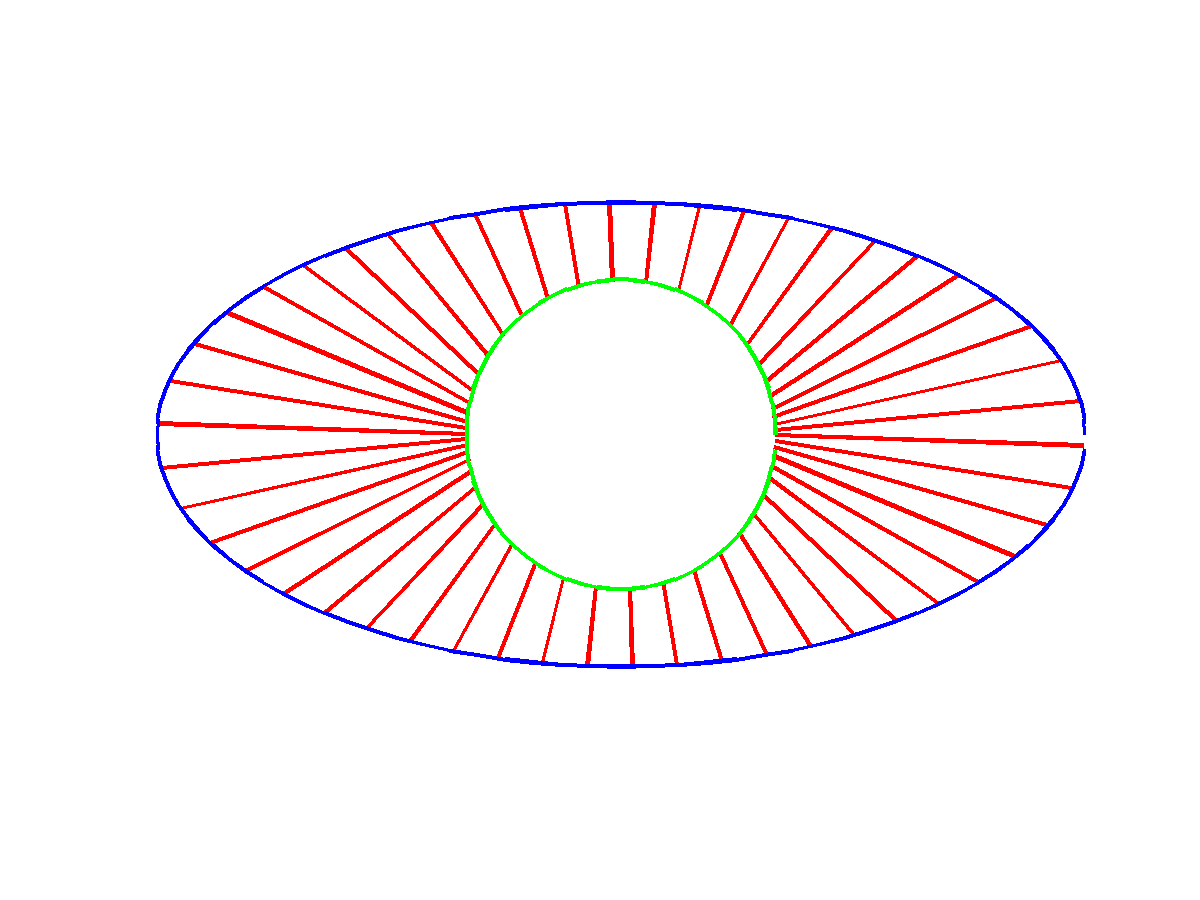}}%
   \subfloat[Transformed diffusion mapping]{\label{fig:circle_trans_diff}%
    \includegraphics[trim=2cm 3cm 1.5cm 1cm, clip=true, width=0.33\textwidth]{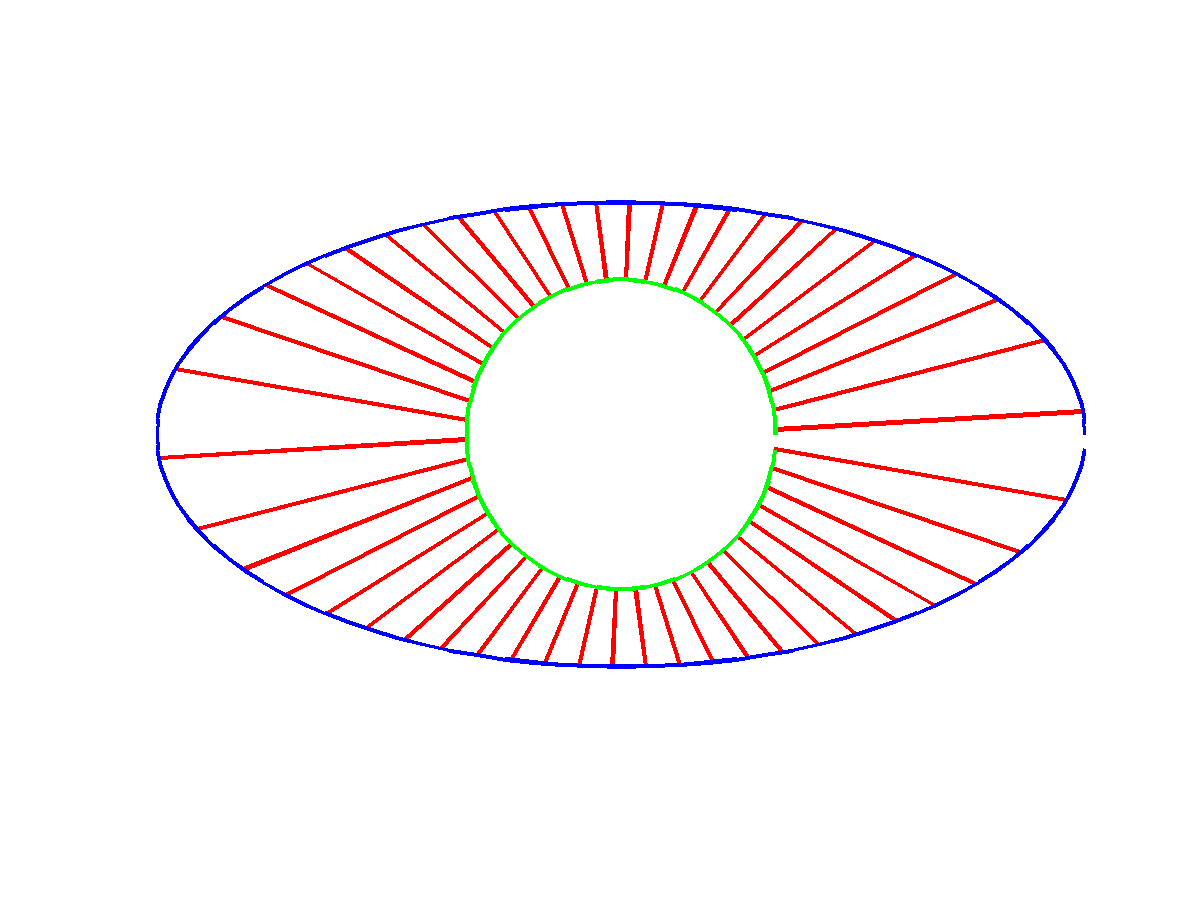}}%
 \quad%
 \quad%
\caption{\textbf{Comparison of Mapping Algorithms with a Simple Test Case.} Displacement fields (red) for the simple mapping problem between $C_1$, a circle (green) and $C_2$, an ellipse (blue), as generated by mapping algorithms based on minimal distance, normal vectors, or diffusion, as well as their reverse and transformed versions.\label{fig:circle_ellipse}}
 \end{center}
\end{figure}

Here we present results of a qualitative evaluation of the basic mapping algorithms. We first consider the simple case of mapping a circle onto a larger ellipse (Fig.\, \ref{fig:circle_ellipse}).

\begin{itemize}
  \item \textit{Minimal Distance: } 
  Mappings based on the minimal distance between two points are easy to implement but prone to errors 
since the algorithm only gives good results if the domain grows uniformly. Problems occur as soon as  big changes in curvature are observed, which are typical for branching morphogenesis and morphogenesis in general. Figure \ref{fig:short_dist_prob} shows that the minimal distance algorithm fails to map a straight line to the line with a protrusion; no point on $C_1$ is mapped onto the curve segment between points A and B. Minimal distance mapping strongly depends on the direction of mapping. In many cases reverse minimal distance mapping provides better results (Fig.\ \ref{fig:rev_short_dist_wins}), because the complexity of the domain structure typically increases rather than decreases during development. Our circle-ellipse example illustrates the limitations of minimal distance mapping (Fig.\,\ref{fig:circle_min}). The problem can be resolved by reversing the algorithm (Fig.\,\ref{fig:circle_rev_min}). A slight improvement is observed if we do a transformation beforehand (Fig. \ref{fig:circle_trans_min}). \\

\begin{figure}[h]   
\begin{center}
  \subfloat[Minimal distance]{\label{fig:short_dist_prob}%
    \includegraphics[width=0.4\textwidth]{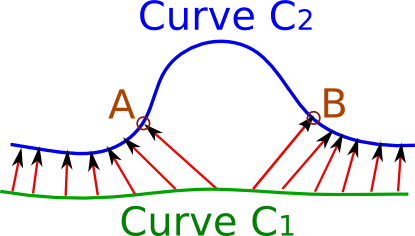}}%
 \quad%
 \subfloat[Reverse minimal distance]{\label{fig:rev_short_dist_wins}%
    \includegraphics[width=0.4\textwidth]{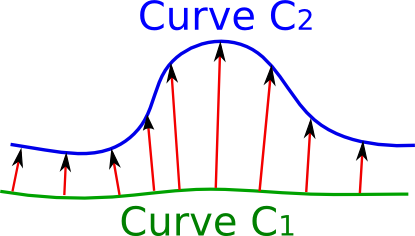}}
 \quad%
 \caption{\textbf{Minimal Distance Mapping.} Every point on the curve $C_1$ is mapped to the point on the curve $C_2$ to which it has the minimal distance. If curve $C_2$ shows big changes in curvature, it is often advisable to use reverse minimal distance mapping. (a) Minimal distance mapping cannot map points onto points in the segment between points A and B. (b) Reverse minimal distance mapping can resolve the issue shown in panel a.}
 \end{center}
  \label{fig:rev_vs_short_dist}
\end{figure}
  \item  \textit{Normal Mapping: } %
  The normal mapping algorithm provides an excellent mapping for the simple test case of a circle and a cylinder (Fig.\,\ref{fig:circle_normal}). Unlike the minimal distance algorithm, the normal mapping algorithm is not very much affected by the curvature of $C_2$. It is, however, sensitive to the curvature of $C_1$.  If the curvature 
in locally concave regions of $C_1$ is comparably higher than the distance between the curves, then crossings of the displacement field vectors can be observed (Fig.\,\ref{fig:normal_problem}). Therefore, the quality of a displacement field computed by normal mapping strongly depends on whether $C_1$ encloses a concave domain and whether $C_2$ is far away from these concave regions. This problem is illustrated in Fig.\,\ref{fig:normal_problem} where the two curves look very much alike but are far away from each other and the enclosed domain of $C_1$ is not convex. This case results in a crossing of the displacement field vectors, which is not desirable. As shown in Fig.\,\ref{fig:normal_problem_solved} this problem is locally solved by doing a reverse mapping.  In case of our simple circle-ellipse example we observe that a normal mapping works very well because a circle is a concave domain (Fig.\,\ref{fig:circle_normal}). On the other a hand, a reverse normal mapping in this configuration (Fig.\,\ref{fig:circle_rev_normal}) leads to crossings of the displacement field vectors and large parts of the ellipse are not mapped onto the circle because the normal vectors from these regions fail to reach the circle. This problem can be addressed by a re-scaling of the domains prior to normal mapping (Fig.\,\ref{fig:circle_trans_normal}), such that similar results are obtained as with the original normal mapping. \\

\begin{figure}[t]   
\begin{center}
  \subfloat[Normal mapping]{\label{fig:normal_problem}%
    \includegraphics[width=0.4\textwidth]{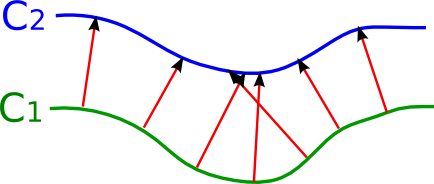}}%
 \quad%
 \subfloat[Reverse normal mapping]{\label{fig:normal_problem_solved}%
    \includegraphics[width=0.4\textwidth]{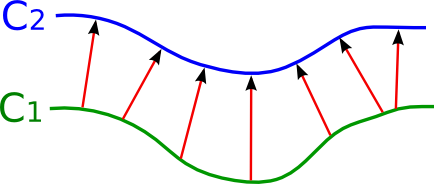}}
 \quad%
 \caption{\textbf{Normal mapping.} (a) If $C_1$ encloses a non--convex domain and both curves are not very close to each other in these concave regions of $C_1$, then the displacement field vectors generated by normal mapping cross each other. (b) The problem can be resolved by using reverse normal mapping.}
 \end{center}
  \label{fig:normal_problem_and_solution}
\end{figure}

\begin{figure}
  \centering
  \subfloat[Normal mapping fails]{\label{fig:normal_and_min_dist_problem1}%
  \includegraphics[width=0.4\textwidth]{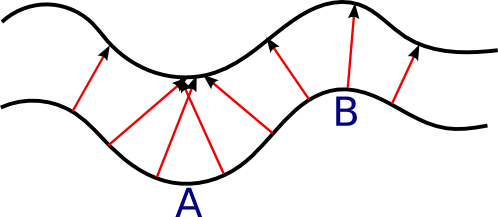}}%
  \quad%
 \subfloat[Minimal distance fails]{\label{fig:normal_and_min_dist_problem2}%
  \includegraphics[width=0.4\textwidth]{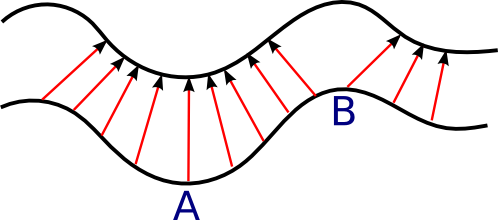}}
  \quad%
  \caption{\textbf{Normal and Minimal Distance Mapping Fail.} When we try to map curves of changing curvature sign that lie far apart from each other, we run into problems with both minimal distance and normal mapping. We have a symmetric problem that cannot be solved by reversing the algorithm. In (a) we do a normal mapping that leads to crossings in region A that will just lead to crossings in region B when we reverse the algorithm, and in (b) minimal distance gives us bad results in region B since we are ignoring some parts of the curve onto which we are mapping to. This can also not be solved by using reverse minimal distance because the same problem would show up in curve region A.}
  \label{fig:normal_problem_and_solution}
\end{figure}

  \item  \textit{Diffusion Mapping: } %
  Many real world mapping problems can be resolved by only mapping curve segments onto each other, because they often only have one curvature direction, such that we can find a good mappings based on normal or minimal distance mapping, in combination with the appropriate prior transformations. However, if the curvature changes its sign within the subdomain, i.e. if the curve changes between being locally convex or concave (Fig. \ref{fig:normal_problem_and_solution}), minimal distance, normal mapping and the reverse mappings all yield bad results in some section of the curve. In Figure\,\ref{fig:normal_and_min_dist_problem1}, normal mapping gives good results in region B where the curve we are mapping from is locally concave and returns crossings in region A where the curve is locally convex. 
  Minimal distance mapping fails in the opposite situations (Fig.\,\ref{fig:normal_and_min_dist_problem2}) and the reverse mapping will help in neither case due to the symmetry of the problem.
Good mapping results can, however, be obtained with diffusion mapping, because the streamlines reach any region of $C_2$ and will never cross each other. 
The downside of the diffusion method is that small regions on the generally shorter curve $C_1$ are mapped onto fairly large regions on $C_2$ as can be seen in Figure\,\ref{fig:circle_diff}. In this case, the reverse mapping can help (Fig.\,\ref{fig:circle_rev_diff}) as well as a first scaling of $C_1$ (Fig.\,\ref{fig:circle_trans_diff}). 

Diffusion mapping is very safe but does not give overall as good results as normal and minimal distance mapping. Furthermore, it is computationally much more expensive than any of the other mapping algorithms and it is therefore wise to only apply it when normal or minimal distance fail, due to their limitations on curves that have changing curvature sign and lie far apart from each other. \\

  \item   {\textit{Uniform Mapping: }} %
  Uniform mapping is the method of choice for open curves with relatively small deformation (Fig.\,\ref{fig:uniform_mapping}). For closed curves or curve segments between two intersection points this method is advisable only if the curves are relatively short or the changes in curvature are small because the direction of the displacement field is always biased towards the direction of the biggest change in growth, as can be seen in Figure\,\ref{fig:uniform_fails}.\\

\begin{figure}[b!]   
\begin{center}
  \subfloat[]{\label{fig:uniform_mapping}%
    \includegraphics[width=0.45\textwidth]{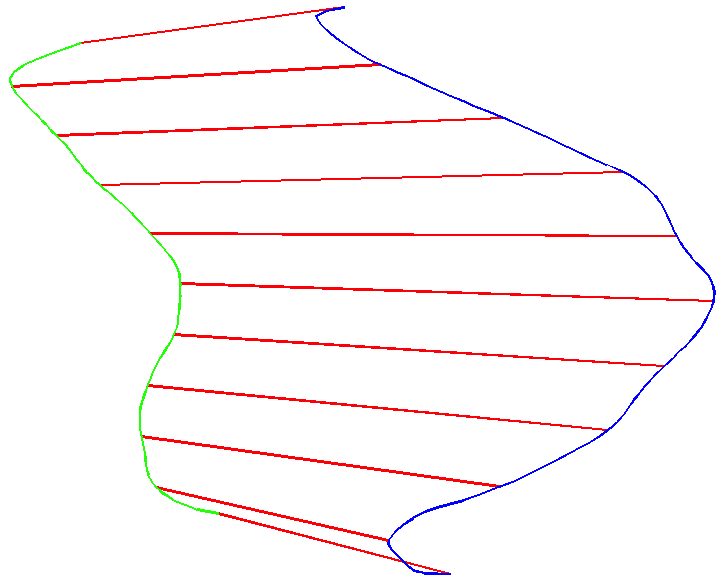}}%
 \quad%
 \subfloat[]{\label{fig:uniform_fails}%
    \includegraphics[width=0.45\textwidth]{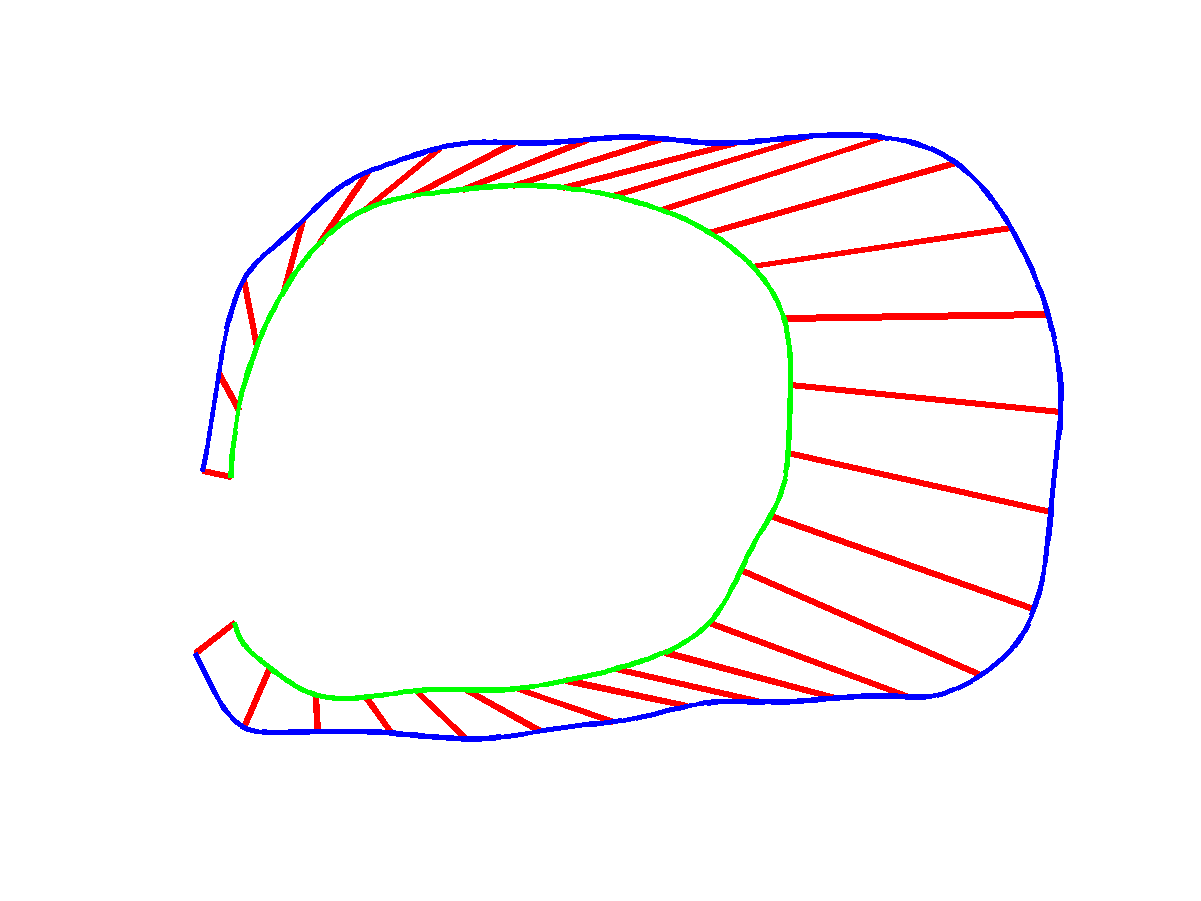}}
 \quad%
 \caption{\textbf{Uniform Mapping} (a) The displacement fields of open curves are determined by interpolating both curves with equidistant points and by mapping the points consecutively onto each other. (b) When the growth of the domain is not uniform, we get a displacement field that is biased towards the direction of maximal growth.}
 \end{center}
\end{figure}

  \item  {\textit{Transformed Mapping: }} %
  We have applied our basic algorithms to the circle-ellipse example (Fig.\,\ref{fig:circle_ellipse}) and saw that scaling the circle beforehand could lead to an improvement of the displacement field quality. Now we apply the transformation algorithm to our mouse kidney data and compare it qualitatively with the non-transformed algorithm (Fig.\,\ref{fig:kidney_no_transform}).
 In Figure\,\ref{fig:kidney_no_transform} we see that minimal distance fails in many curve regions. The transformation algorithm first scales and aligns the curve $C_1$ which results in the dashed curve $C_{1,t}$ (Fig.\,\ref{fig:kidney_scaled}). Then $C_{1,t}$ is mapped onto $C_2$ (Fig.\,\ref{fig:kidney_during_transform}). The starting points of the displacement field vectors that lie on $C_{1,t}$, are back transformed onto $C_1$ such that we get the final displacement field vectors (Fig.\,\ref{fig:kidney_with_transform}). We observe that the newly obtained displacement field has a much higher quality than without prior transformation since $C_{1,t}$ and $C_2$ are on average closer to each other than $C_1$ to $C_2$. The downside of this method is that the displacement field vectors are very often distorted, i.e. the angles between the vectors and the $C_1$ normal vectors are large.
When do we want to first scale the curve $C_1$? The more uniformly the growing domain develops, the better it is to scale $C_1$ prior to doing the mapping. 
Usually there are many curve segments  $C_{in} \subseteq C_2$ that lie inside the domain bounded by $C_1$. By scaling, we usually improve the mapping from $C_1$ to $C_{out} = C_2 \backslash C_{in}$ and worsen the mapping results from $C_1$ to $C_{in}$. So it is not advisable to do scaling if $C_{in}$ is large and does not always lie close to $C_1$. This is often the case if we map $C_1$ onto a much later developed domain $C_2$ since the complexity of the curves are increasing and do not only grow uniformly.

\begin{figure}[t]   
\begin{center}
  \subfloat[Minimal distance mapping]{\label{fig:kidney_no_transform}%
    \includegraphics[trim=3.7cm 2cm 2cm 0.5cm, clip=true, width=0.45\textwidth]{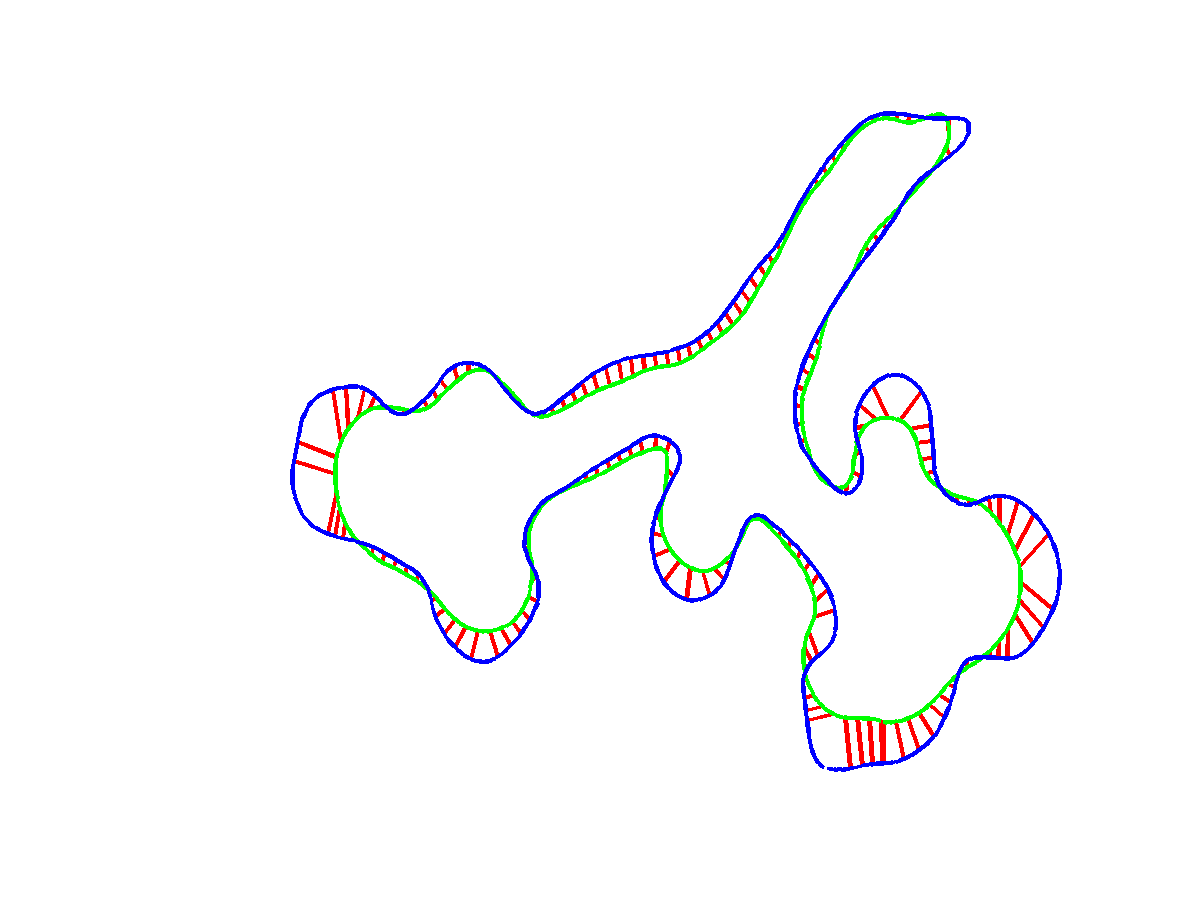}}%
 \quad%
 \subfloat[$C_1$ scaled indicated by dashed line]{\label{fig:kidney_scaled}%
    \includegraphics[trim=3.7cm 2cm 2cm 0.5cm, clip=true, width=0.45\textwidth]{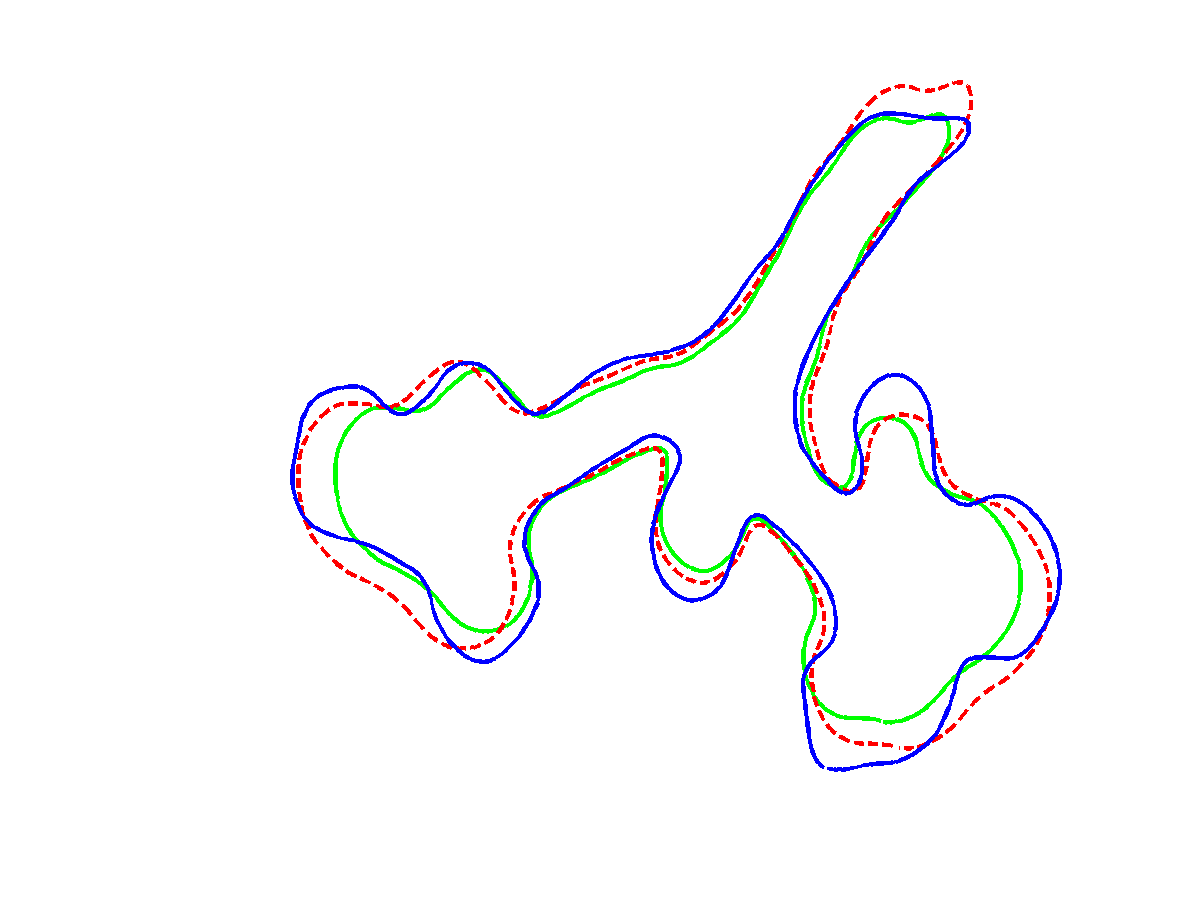}}
 \quad%
   \subfloat[Minimal distance mapping from $C_{\text{t}}$ onto $C_2$]{\label{fig:kidney_during_transform}%
    \includegraphics[trim=3.7cm 2cm 2cm 0.5cm, clip=true, width=0.45\textwidth]{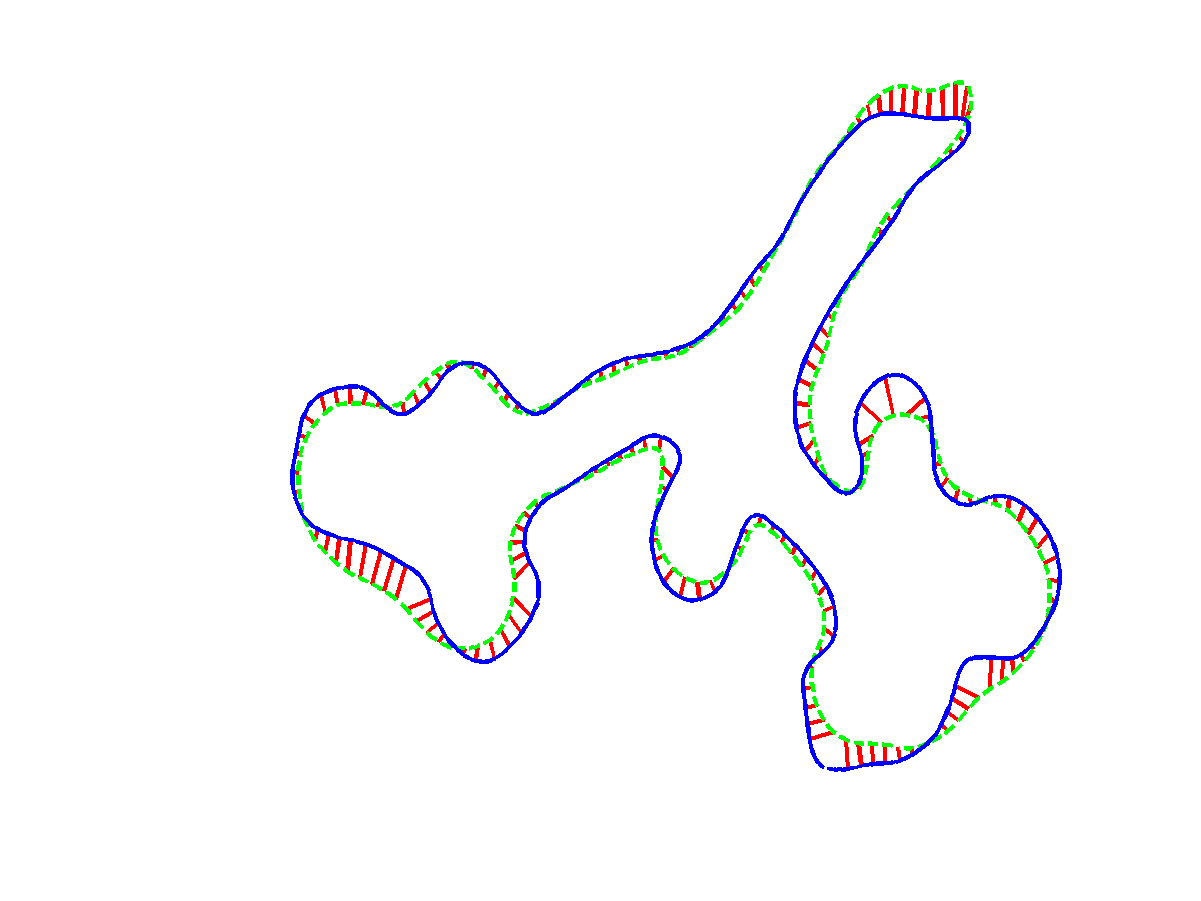}}%
 \quad%
 \subfloat[Displacement field starting points are transformed back onto $C_1$]{\label{fig:kidney_with_transform}%
    \includegraphics[trim=3.7cm 2cm 2cm 0.5cm, clip=true, width=0.45\textwidth]{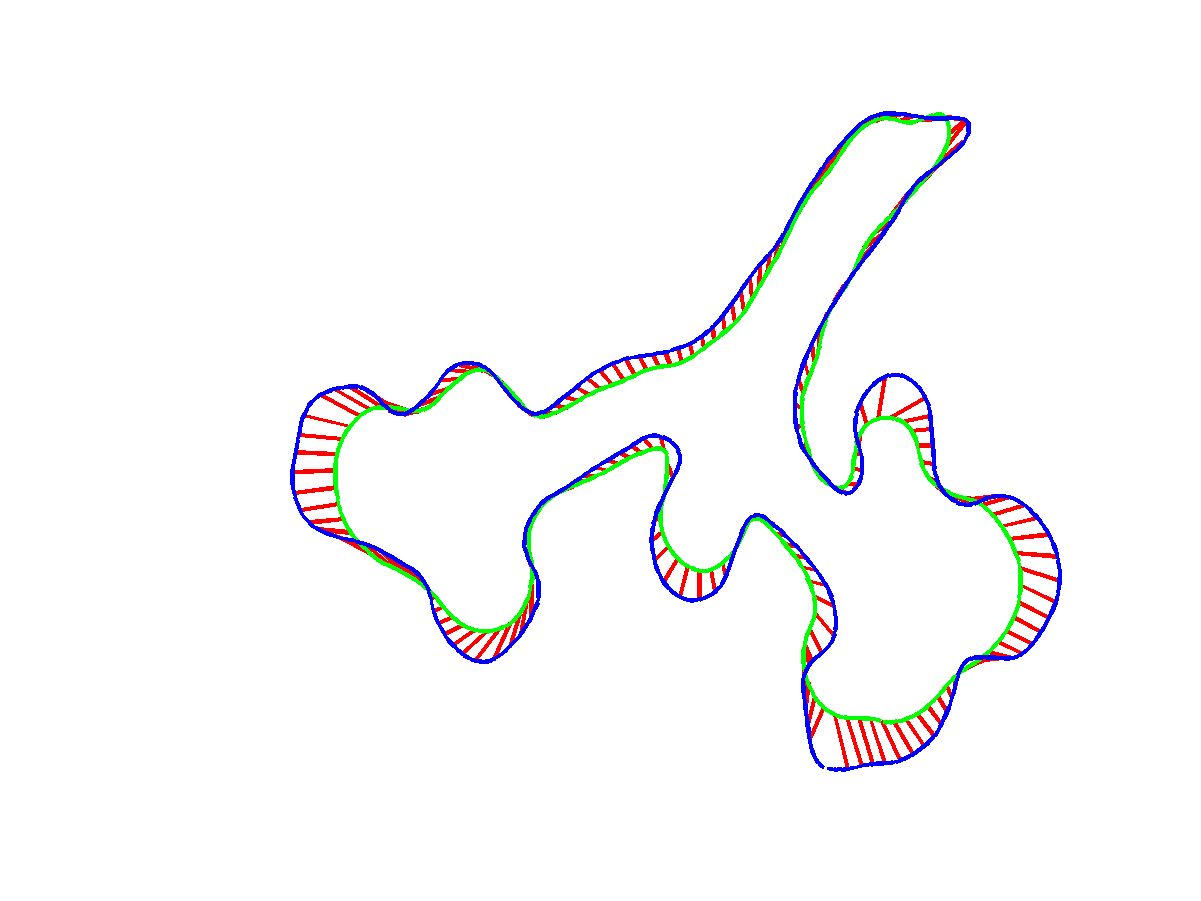}}
 \quad%
 \caption{\textbf{Transformation Prior Mapping.} (a) The displacement field was computed using the basic minimal distance algorithm. (b) By scaling the inner curve $C_1$ onto the same size as $C_2$ the two curves got closer to each other. (c) We computed the displacement field from our transformed curve $C_{1,t}$ onto $C_2$. (d) The starting points of our vectors where then transformed back onto our original curve $C_1$. We observe that in (d) the number of regions on $C_2$, onto which no points where mapped to, has decreased and the displacement field quality has therefore increased.}
 \end{center}
  \label{fig:transformation_kidney}
\end{figure}

\end{itemize}

In summary, we showed that the different algorithms are suited for different mapping problems. The normal and minimal distance mapping algorithms often produce the nicest displacement fields, but they fail for curves with strong changes in the curvature. The diffusion mapping algorithm rarely fails, but overall gives sub-optimal results.

\subsection{ {Mapping in the Limiting Case of an Infinite Number of Points}}

In this section we will discuss the mapping algorithms for smooth curves and the consequences in a discrete space. In case of smooth curves, the displacement field algorithm $\mathcal{D}$: $C_1 \rightarrow C_2$, is applied onto all the infinite number of points along the curve $C_1$ that are mapped onto $C_2$. We call the mapping  well-defined on a subset $\tilde{C}_1 \subseteq C_1$ if \space $ \forall p_1\in \tilde{C}_1, \exists p_2 = \mathcal{D}(p_1) \in C_2$. Both uniform and diffusion mapping have the properties that $\tilde{C}_1 = C_1$ and the mapping is bijective. These are desirable properties since a deforming domain will neither have boundary points that vanish nor will points contract to a single point. Furthermore, both algorithms have the property that they do not allow crossings of the theoretical displacement field. When we talk about the \textsl{crossing} of the  displacement field, we mean that each point on $C_1$ is mapped onto $C_2$ such that the order in which the points occur along the curve is conserved.

Let us now look at minimal distance and normal mapping and the reverse of both of them. The minimal distance algorithm applied to smooth curves will always return results that are normal to the curve on which they are mapped onto. Therefore, reverse minimal distance and normal mapping as well as minimal distance and reverse normal mapping are intrinsically the same with the only difference that (reverse) minimal distance on a closed and smooth curve only returns a subset of the displacement field that we get by (reverse) normal mapping. Without loss of generality we will only look at the first algorithm pair: normal and reverse minimal distance mapping. 

Since the minimal distance algorithm is well-defined on $C_1$, the reverse minimal distance algorithm maps every point $p_2 \in C_2$ onto a point $p_1 \in C_1$. This means that we find a mapping from a subset of points $\tilde{C}_1$ to all of $C_2$. Therefore, if a continuous part on $C_1\backslash \tilde{C}_1$ is mapped onto $C_2$ using normal mapping, there must be crossings in the displacement field except if all points of $C_1\backslash \tilde{C}_1$ map to exactly the same point, i.\,e. they lie on a perfect circle and map to its centre. By this we see that reverse minimal distance discards all the mapping vectors that could lead to crossings. Note that this property no longer holds in regions close to the boundary of two open curves of finite length because minimal distance might map points onto the start and end points of $C_2$ and these vectors will no longer be normal to the curve.

For our computations it is important to now consider the case where we have a finite number of points that are mapped onto each other.

Despite our considerations for the mapping algorithms for smooth curves from above, we mostly observe that normal mapping provides better results than reverse minimal distance for the following three reasons: First of all, if the density of the points that should be mapped is not very high, there will be no crossings for normal mapping and reverse minimal distance would still have its usual problems of missing out some regions on $C_1$. Secondly, reverse minimal distance yields bad results on the boundary of open curves.  {Thirdly, in generalisation of the second point, depending on our interpolation method, e.g. when using linear interpolation, our curve will not be smooth.} As a result, the displacement field vectors will no longer be perfectly normal on $C_1$ and reverse minimal distance will preferably map onto local extrema that are given by the points where $C_1$ is not smooth. Therefore, if we map closed curves onto each other, it is advisable to first compute the density of the points and do reverse minimal distance mapping if the density exceeds some given density threshold; otherwise we use normal mapping. 

\subsection{ {Quantitative Evaluation of the Mapping Algorithms}} 
 \label{criterias}

\begin{figure}[b!]
\begin{center}
\includegraphics[width=0.7\textwidth]{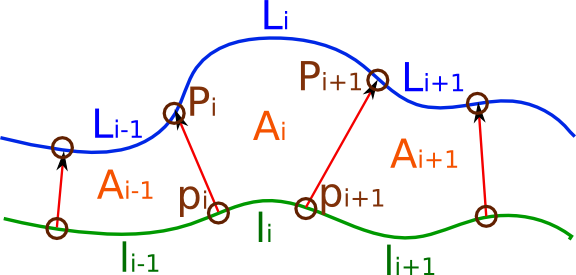}
\caption{\textbf{Quantifying the Uniformity of the Mapping.} The quality measure (Eq.\,\ref{Eq_QM}) is based on the segment lengths $L_i$ and $l_i$ between the two mapped points $P_i$ and $P_{i+1}$ and the original points $p_i$ and $p_{i+1}$ respectively. The enclosed surface is defined as $A_i$.}
\label{fig:qm_notation}
\end{center}
\end{figure}

We now seek  to define a measure that quantifies the quality of the displacement field. An important criterium is the uniformity of the mapping: a good mapping should map a small arc length on $C_1$ also onto a small arc length of $C_2$, which leads us to the quality measure
\begin{equation}\label{Eq_QM}
 \text{QM}_1 = \left < \left (\frac{L_i}{\alpha  l_i} - 1 \right)^2  \right >.
\end{equation}
Here, $l_i$ is the arc length on $C_1$ between the points $p_i$ and $p_{i+1}$, which are mapped onto the points $P_i$ and $P_{i+1}$ on $C_2$ with arc length $L_i$ between them, as shown in Figure \ref{fig:qm_notation}. The value $\alpha$ is defined as the the scaling factor such that
\begin{equation}
\alpha = \text{Circumference}(C_2)/\text{Circumference}(C_1) 
\end{equation}
and $<\cdot>$ is the average over all $i$. The smaller the value QM$_1$, the better the displacement field. If all segments $l_\mathrm{i}$ and $L_\mathrm{i}$ on curves $C_1$ and $C_2$ respectively  have equal length, we have the minimal value, QM$_1=0$. This is naturally the case for uniform mapping.

An alternative measure for the uniformity of the displacement field can be formulated based on the area $A_\mathrm{i}$ enclosed by the segments $l_\mathrm{i}$, $L_\mathrm{i}$ and mapping vectors (Fig.\,\ref{fig:qm_notation}):

\begin{equation}
\text{QM}_2 = \left < \left(\frac{A_i}{<A_i>}-1\right)^2 \right >
\end{equation}
where $A_i$ is the area defined by the two arc lengths and the displacement vectors.  Finally we can combine the two measures, QM$_1$ and QM$_2$, to obtain:
\begin{equation}
\text{QM} = \text{QM}_1 \cdot \text{QM}_2 =  \left < \left(\frac{L_i}{\alpha  l_i} - 1\right)^2  \right > \cdot  \left <\left(\frac{A_i}{<A_i>}-1\right)^2 \right > 
\end{equation}\\

Table \ref{qm_results} shows the calculated values of QM$_1$, QM$_2$ and QM for the simple test case of the circle being mapped onto the ellipse (Fig.\,\ref{fig:circle_ellipse}). The three measures yield the same ranking within the basic algorithms for minimal distance and normal mappings. However, in case of diffusion mapping, QM$_1$ favours the transformed mapping while QM$_2$ favours the reverse mapping. The combined measure QM yields the same ranking as QM$_1$. Visual inspection confirms that all  highlighted mappings deliver excellent results (Fig.\,\ref{fig:circle_ellipse}), with QM$_1$ and QM delivering the most appropriate ranking. \\

 \begin{table}[t]
\caption{Quality measures calculated for the test case of a circle mapped onto an ellipse (Fig.\,\ref{fig:circle_ellipse}). The lowest values for the quality measures are marked in bold. \label{qm_results}}
\begin{center}
\begin{tabular}{c | c | c | c } 
\hline 
Mapping & QM$_1$ & QM$_2$ & QM \\[0.5ex] \hline 
minimal distance  mapping                      & 13.4 & 22.4  & 301  \\
reverse  minimal distance        & \textbf{0.128} & \textbf{0.134} & \textbf{0.017}  \\
transformed minimal distance   &  1.69 & 2.71 & 4.58 \\ \hline
normal mapping                                          & 0.103 & 0.398 & 0.041 \\
reverse  normal                         & 56.1 & 2.45 & 137 \\
transformed normal                    & \textbf{0.084} & \textbf{0.397} & \textbf{0.033} \\ \hline
diffusion mapping                                       & 0.494 & 1.17 & 0.577 \\
reverse  diffusion                      & 1.589 & \textbf{0.088} & 0.140 \\
transformed diffusion                 & \textbf{0.224} & 0.559 & \textbf{0.125} \\ [1ex] 
\hline 
\end{tabular}
\end{center}
\end{table}

\begin{figure}[h]   
\begin{center}
  \subfloat[Minimal distance mapping]{\label{fig:kid_min}%
    \includegraphics[trim=2cm 2.5cm 1.5cm 2cm, clip=true, width=0.43\textwidth]{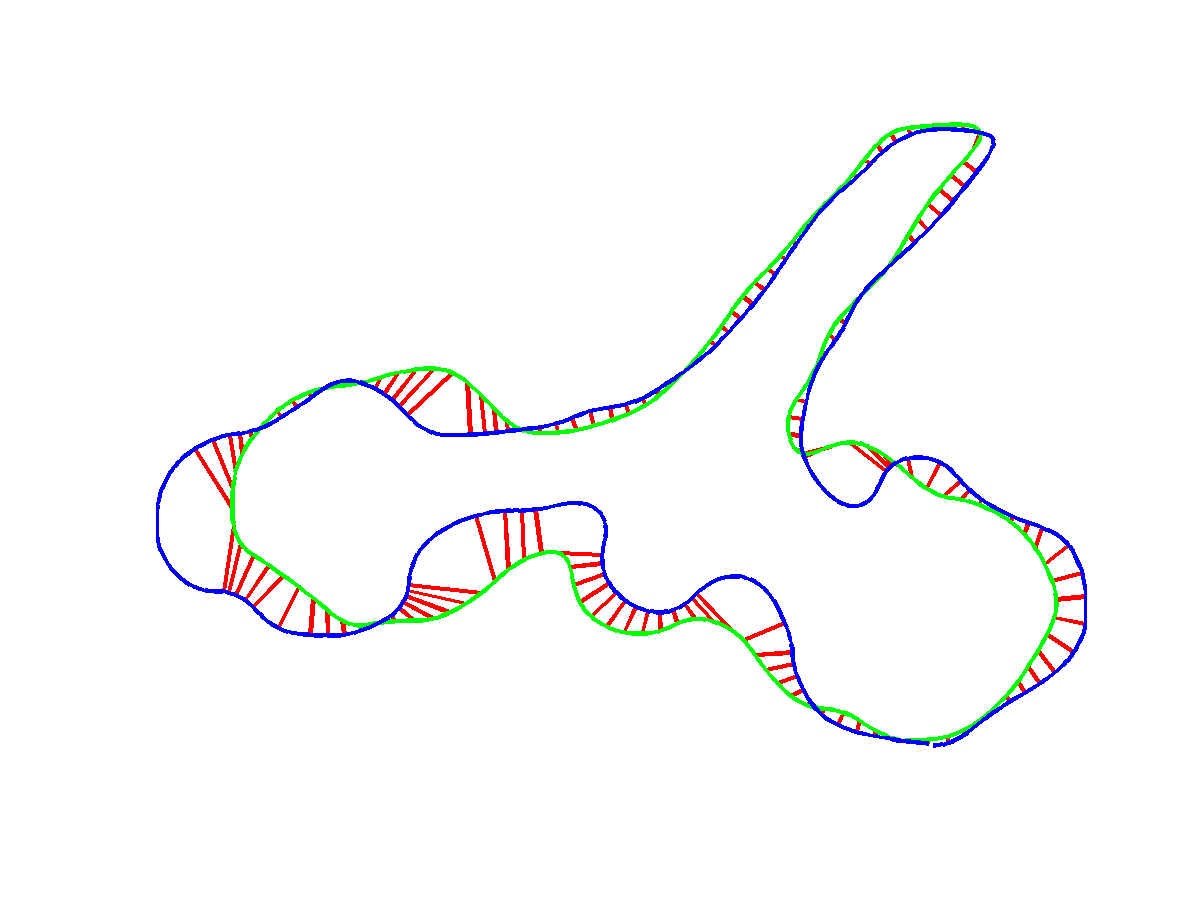}}%
   \subfloat[Reverse minimal distance mapping]{\label{fig:kid_rev_min}%
    \includegraphics[trim=2cm 2.5cm 1.5cm 2cm, clip=true, width=0.43\textwidth]{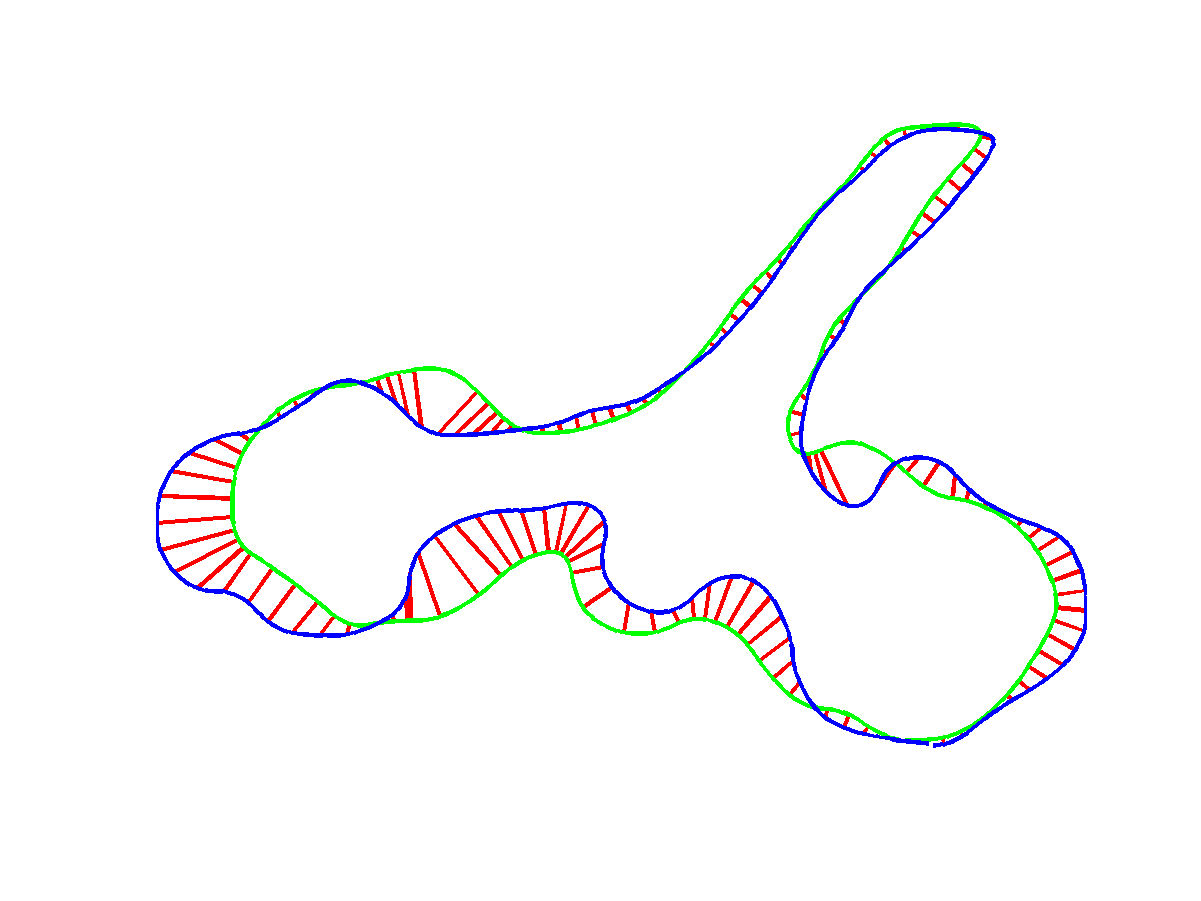}}%
 \quad%
 \subfloat[Normal mapping]{\label{fig:kid_normal}%
    \includegraphics[trim=2cm 2.5cm 1.5cm 2cm, clip=true, width=0.43\textwidth]{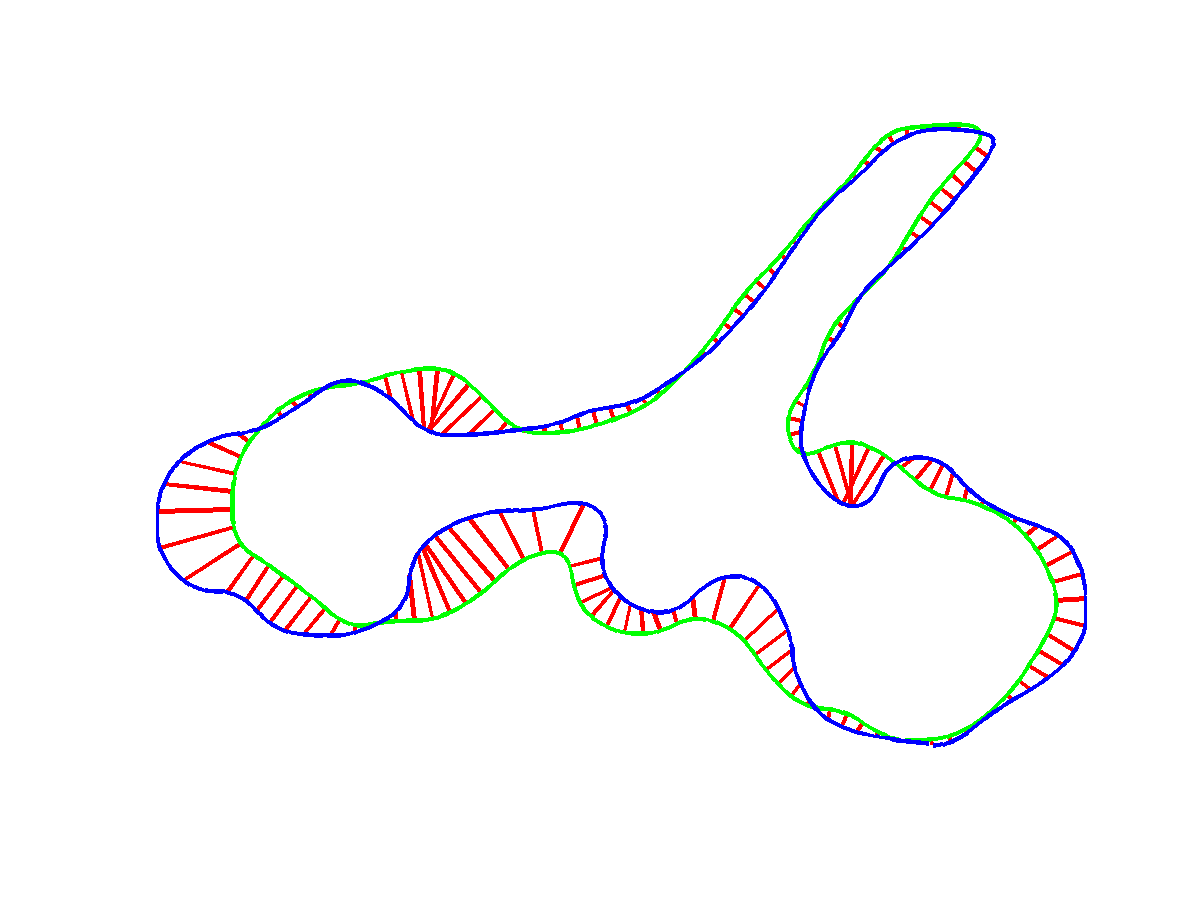}}
  \subfloat[Reverse normal mapping]{\label{fig:kid_rev_normal}%
    \includegraphics[trim=2cm 2.5cm 1.5cm 2cm, clip=true, width=0.43\textwidth]{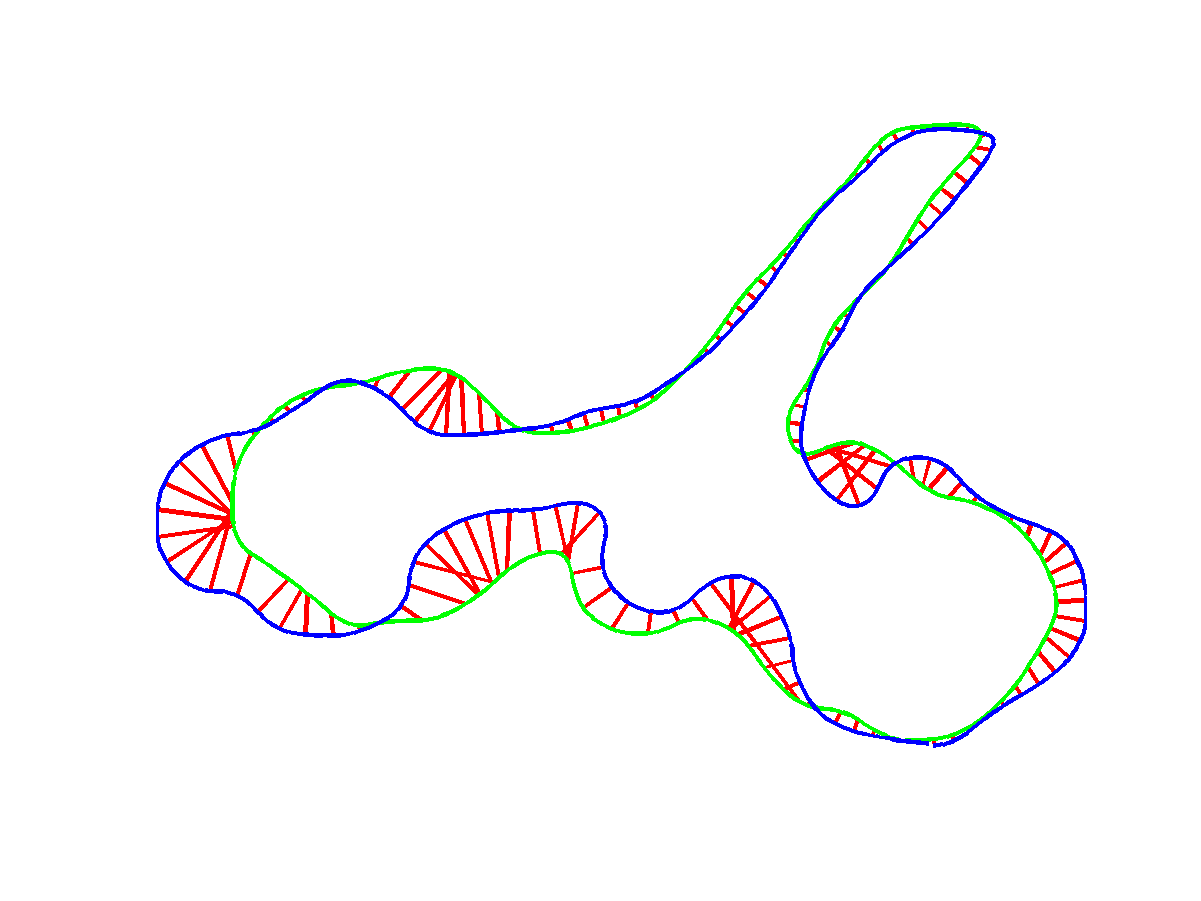}}
 \quad%
 \subfloat[Diffusion mapping]{\label{fig:kid_diff}%
    \includegraphics[trim=2cm 2.5cm 1.5cm 2cm, clip=true, width=0.43\textwidth]{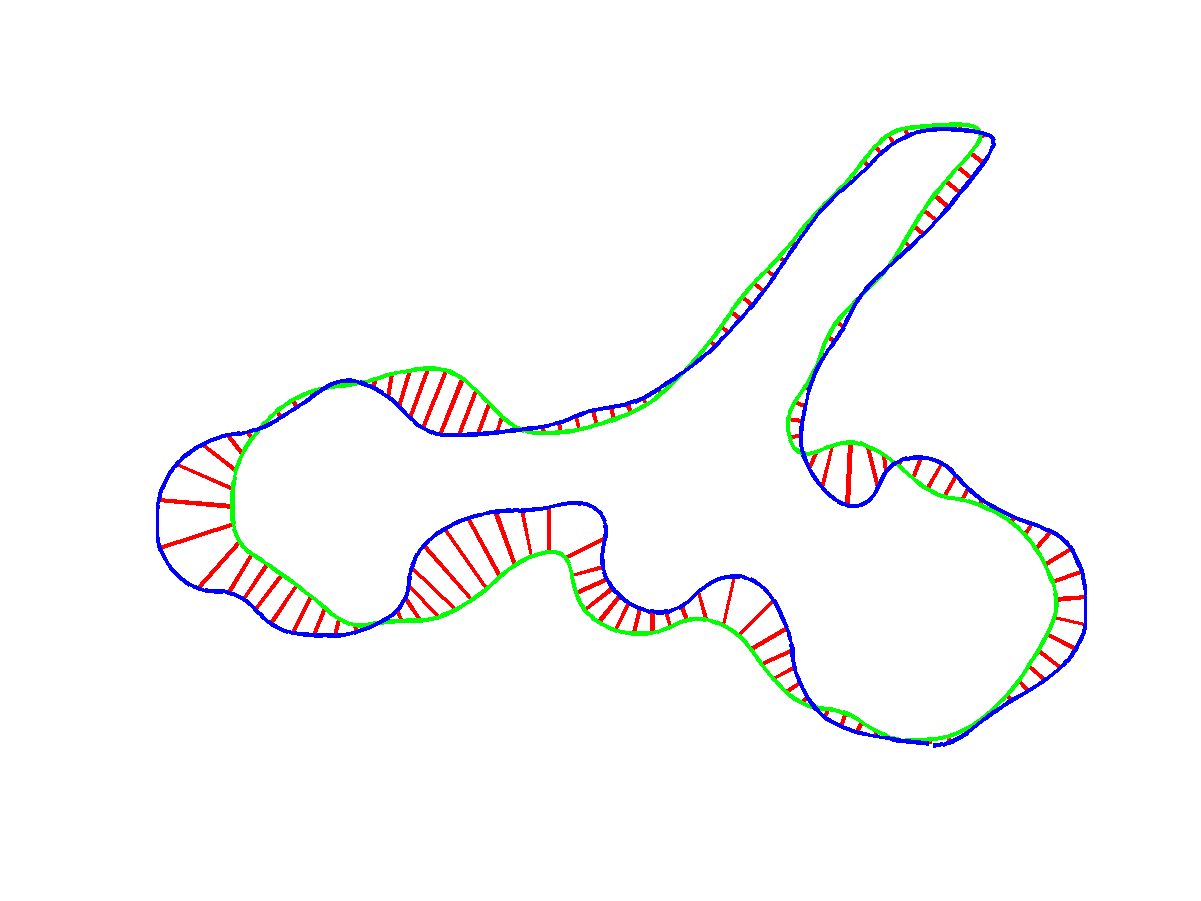}}
   \subfloat[Reverse diffusion mapping]{\label{fig:kid_rev_diff}%
    \includegraphics[trim=2cm 2.5cm 1.5cm 2cm, clip=true, width=0.43\textwidth]{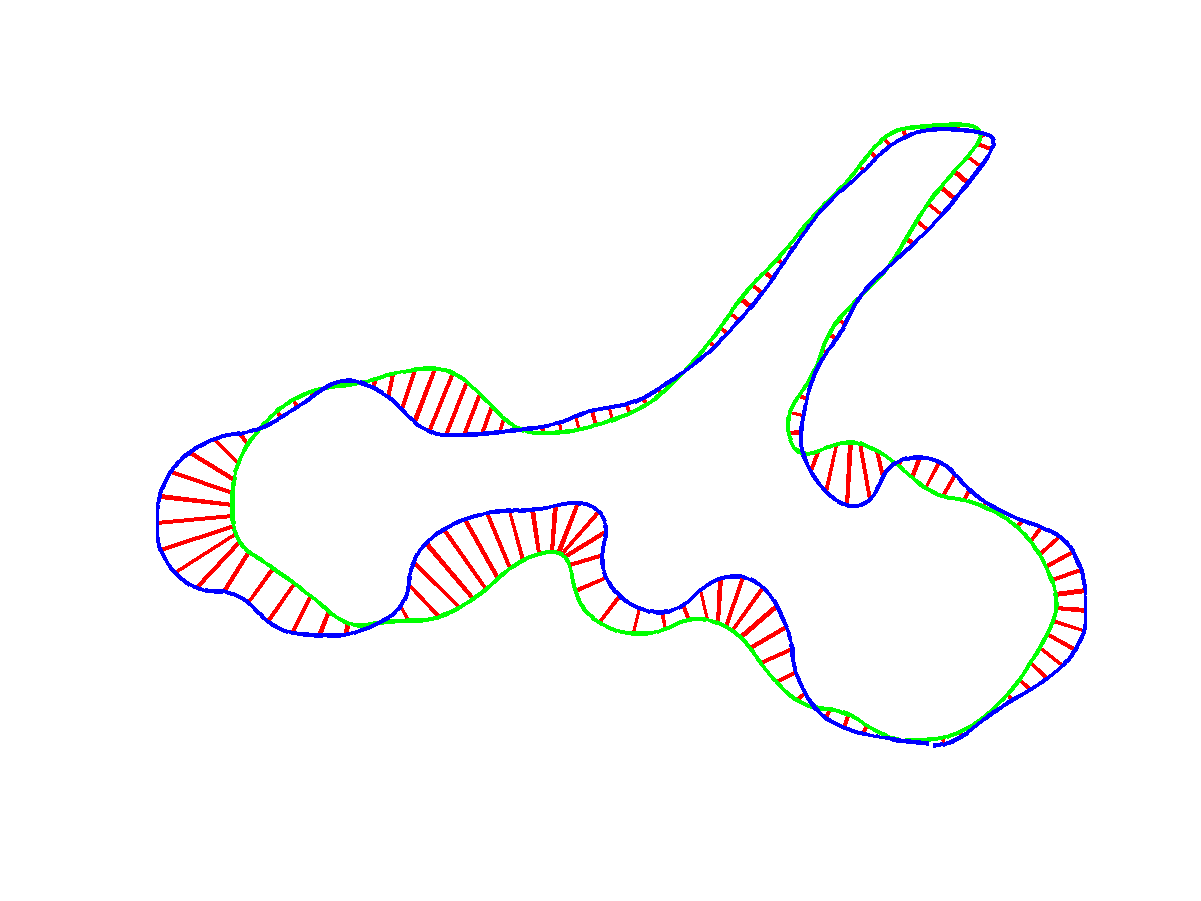}}%
 \quad%
    \subfloat[Uniform mapping]{\label{fig:kid_uni}%
    \includegraphics[trim=2cm 2.5cm 1.5cm 2cm, clip=true, width=0.43\textwidth]{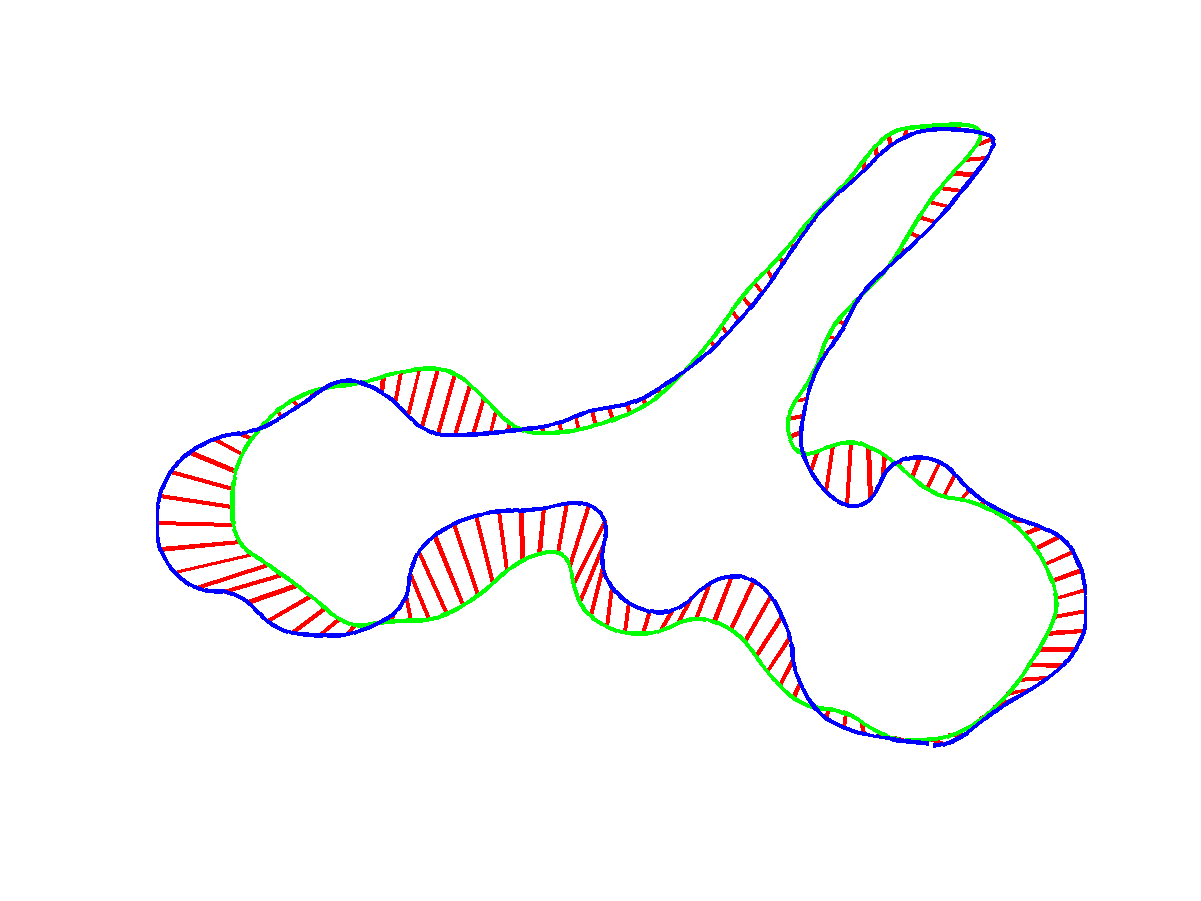}}%
 \quad%
\caption{\textbf{Mapping of Embryonic Dataset.} Displacement fields generated by different methods for our embryonic dataset of a kidney: $C_1$ and $C_2$ represent the shapes of the kidney explant at an initial moment and after 8 hours of development \emph{in vitro}. In all cases we have split our curve into segments before the algorithms where applied.\label{fig:kid}}
 \end{center}
\end{figure}

We repeated the analysis with the dataset for \emph{ex vivo} kidney branching morphogensis \cite{Adivarahan:2013iz}. We used five datasets and mapped the starting embryonic shape to embryonic datasets that had been acquired after 1h, 2h, 4h, 8h (Table \ref{qm_results2}). As expected, the difference between shapes after 1h of development is so small that the quality measures yield more or less the same results; all our mapping algorithms return good and very similar results. The longer the time difference between the two datasets, the worse the mapping, i.e. the value of our quality measures increases as we map curves that are increasingly distant (Table \ref{qm_results2}).  Minimal distance (Fig.\,\ref{fig:kid_min}) and reverse minimal distance (Fig.\,\ref{fig:kid_rev_min}) yield rather poor results for big time steps and strong changes in curvature. The same holds for normal and reverse normal mapping where we observe crossings (Fig.\,\ref{fig:kid_normal} and Fig.\,\ref{fig:kid_rev_normal}); as expected there are more crossings for the reverse normal mapping since it maps from a curve with higher changes in curvature to a curve with lower changes in curvature. The poor results of the normal mapping algorithm are in line with our previous observation that normal mapping gives bad results for very big time steps. The algorithm, however, still yields good results for smaller deformations, i.e. until $\Delta \leq 4$h (Table \ref{qm_results2}).

Uniform mapping naturally gives the best results for QM$_1$ as  it lets QM$_1$ converge to zero by its definition. However, visual inspection reveals that  uniform mapping results in slightly distorted displacement fields (Fig.\,\ref{fig:kid_uni}) compared to the reverse diffusion mapping (Fig.\,\ref{fig:kid_rev_diff}). We note that the solution differs substantially from the one obtained with a normal mapping, i.e.\ the angle between displacement field vectors and the normal vectors are rather large. This is also reflected in QM$_2$, which identifies the reverse diffusion mapping as the best mapping. 

Finally, if we look at QM in Table \ref{qm_results2}, we observe that uniform mapping will again give the best results, since it is almost zero for QM$_1$. However given the distortions that we get for the uniform mapping, uniform mapping is not necessarily the method of choice. Based on the QM value, normal mappings yield the second best result for all cases. The QM ranking does not penalize for crossings. In case of crossings, reverse diffusion would be the method of choice. These considerations are taken into account for our final algorithm for the image-based determination of displacement fields.

\begin{table}[h]
\caption{Quality measure calculated for the test case of the embryonic kidney branching morphogenesis. The lowest values for the quality measures are marked in bold; the uniform mapping was not considered as it gives the lowest value for QM$_1$ and QM by definition.} \label{qm_results2}
\begin{center}
\begin{tabular}{c | c | c | c | c} 
\hline 
Mapping & QM$_1$($\Delta$=1h) & QM$_1$($\Delta$=2h) & QM$_1$($\Delta$=4h) & QM$_1$($\Delta$=8h)  \\[0.5ex] \hline 
minimal distance                & 0.027  & 0.056 &0.294  & 2.03 \\
reverse minimal distance & 0.033  & 0.046 & 0.144  &  71.6\\
normal                                  & \textbf{0.010} & \textbf{0.031}  & \textbf{0.089} & \textbf{0.281} \\
reverse normal                   & 0.154 & 0.108  & 172 &  148.9\\
diffusion                               & 0.019 & 0.041  & 0.193 &  2.03\\
reverse diffusion                & 0.077 &  0.039 & 0.131 &  0.559 \\
uniform mapping               & 0.002 & 0.011  & 0.010 & 0.028\\[1ex]
\hline
       & QM$_2$($\Delta$=1h) & QM$_2$($\Delta$=2h) & QM$_2$($\Delta$=4h) & QM$_2$($\Delta$=8h)  \\[0.5ex] \hline 
minimal distance                & 0.650  &  0.809 & 1.89 & 5.75  \\
reverse minimal distance & 0.587  & 0.593  & 1.009 &  1.748 \\
normal                                  & 0.591 & 0.657  &  1.04 &  1.08\\
reverse normal                   & 0.564 & 0.531  & 0.830 &  1.08\\
diffusion                               & 0.612 & 0.833 & 2.080 &  6.00\\
reverse diffusion                & \textbf{0.506} & \textbf{0.510} & \textbf{0.795} &  \textbf{0.811}\\
uniform mapping               & 0.599 & 0.612 & 0.951  &  0.999\\[1ex]
\hline 
       & QM($\Delta$=1h) & QM($\Delta$=2h) & QM($\Delta$=4h) & QM($\Delta$=8h)  \\[0.5ex] \hline 
minimal distance                & 1.76E-2  &  4.53E-2 & 0.556 & 11.66  \\
reverse minimal distance & 1.94E-2  & 2.73E-2  & 0.145 &  125 \\
normal                                  & \textbf{0.591E-2} & 2.04E-2  &  \textbf{9.25E-2} &  \textbf{0.304}\\
reverse normal                   & 8.69E-2 & 5.73E-2 & 143 &  161\\
diffusion                               & 1.16E-2 & 3.42E-2 & 0.401 &  12.2\\
reverse diffusion                & 3.90E-2 & \textbf{1.99E-2} & 0.104 & 0.454 \\
uniform mapping               & {0.120E-2} & {0.673E-2} & {0.951E-2}  & {0.280}\\[1ex]
\hline 

\end{tabular}
\end{center}
\end{table}

\section{Displacement Fields for 3D Objects}

\begin{figure}
\begin{center}
\includegraphics[width=0.8\textwidth]{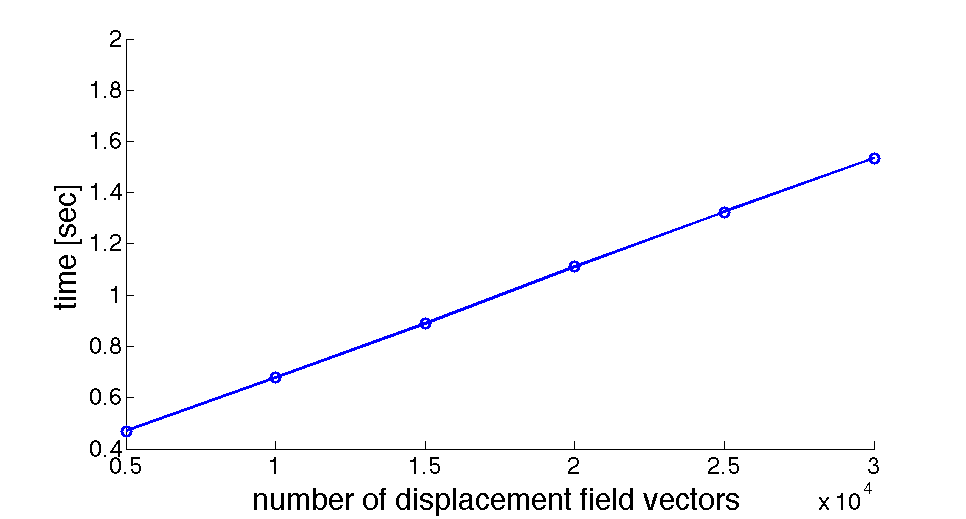} 
\caption{ {\textbf{Benchmark for 3D Normal Mapping.} Using a mesh size of 3'527 and 4'064 mesh-faces, the computational time increases linearly with the number of computed displacement field vectors.}}
\label{fig:benchmark}
\end{center}
\end{figure}

In a final step, we extended the basic algorithms to 3D to enable also 3D simulations on growing domains. The only basic algorithm that cannot be used in 3D is uniform mapping. Since computations in 3D can get very expensive, we have used the OPCODE collision detection library \cite{opcode} for our normal mapping algorithm. This library efficiently computes ray-mash intersection points by creating a collision tree for the possibly ray intersecting mesh.   {To determine the computational efficiency of the algorithm, we determined the computational time that is needed to compute an increasing number of displacement field vectors using our normal mapping algorithm for a mesh size of 3'527 and 4'064 mesh-faces of the 3D surface (Fig.\,\ref{fig:benchmark}). As can be seen, the computational time increases linearly with the number of computed displacement field vectors. }

\begin{figure}
\begin{center}
  \subfloat[Minimal distance mapping]{\label{fig:3d_minimal_distance}%
\includegraphics[trim=3.2cm 3cm 3.2cm 3cm, clip=true, width=0.4\textwidth]{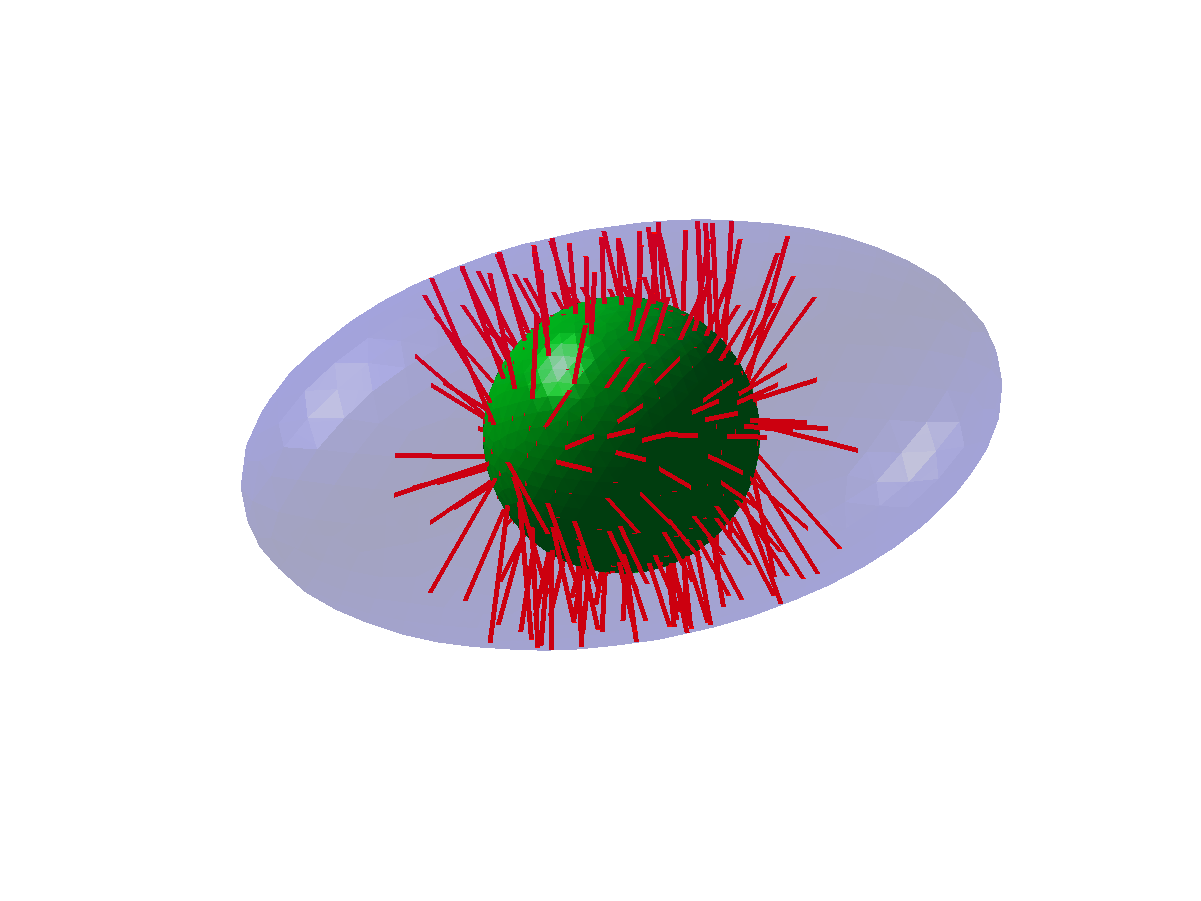}}
\quad%
  \subfloat[Normal mapping]{\label{fig:3d_normal_mapping}%
    \includegraphics[trim=3.2cm 3cm 3.2cm 3cm, clip=true, width=0.4\textwidth]{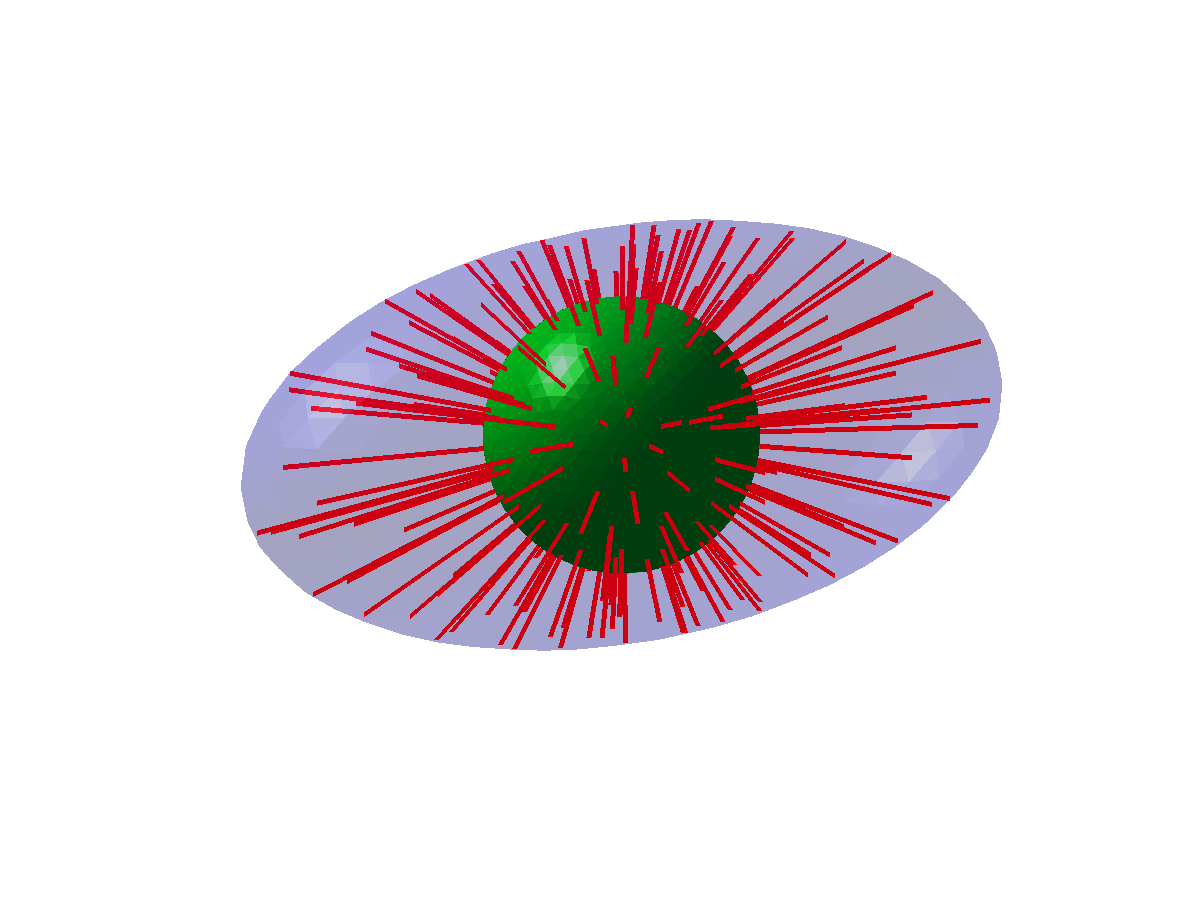}}
 \quad%
    \subfloat[Diffusion mapping]{\label{fig:3d_diffusion_mapping}%
    \includegraphics[trim=3.2cm 3cm 3.2cm 3cm, clip=true, width=0.4\textwidth]{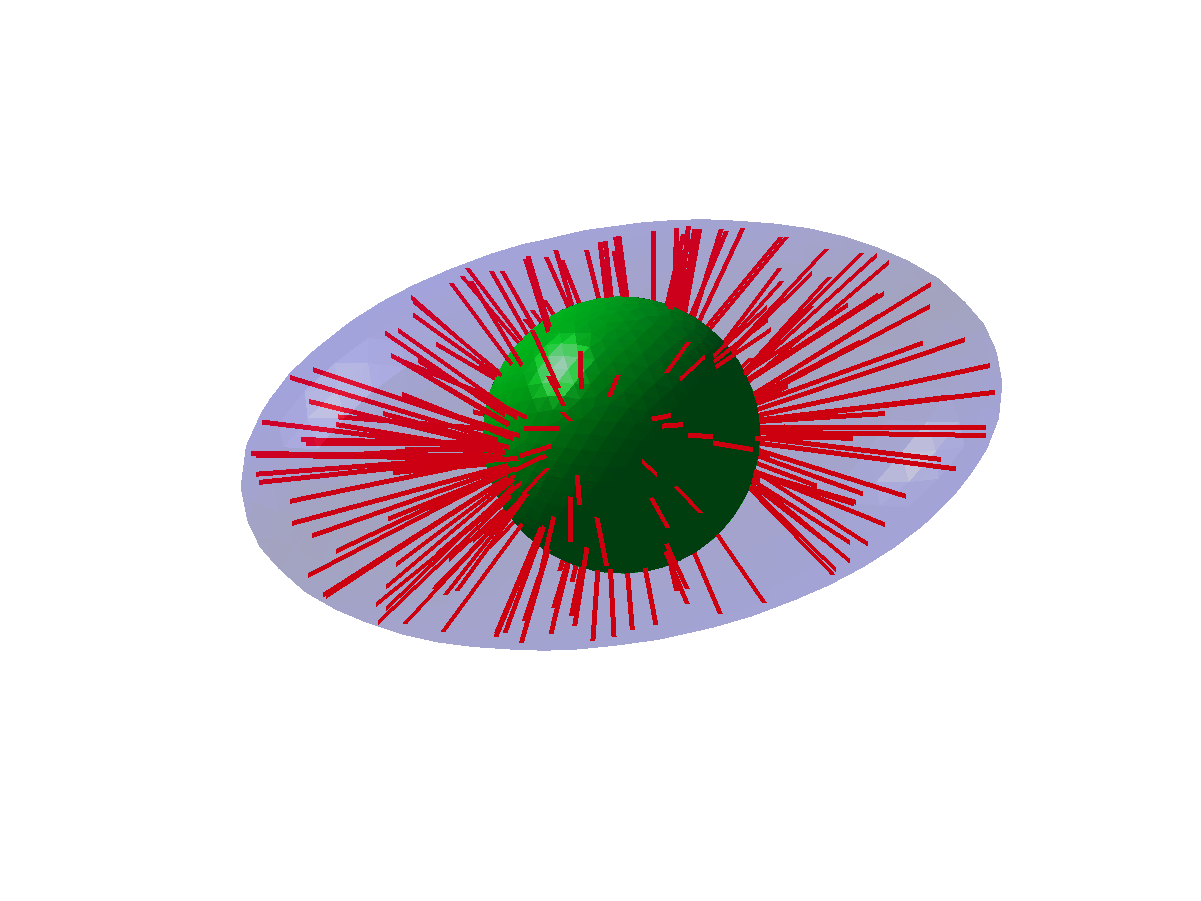}}
 \quad%
\caption{\textbf{3D Minimal Distance and Normal Mapping.} The displacement field from a sphere onto an ellipsoid is computed using (a) minimal distance, (b) normal mapping and (c) diffusion mapping. We observe in (a) that some regions on the ellipsoid cannot be mapped onto and that we get better results in (b) and (c).\label{fig:3d_sphere_ellipsoid}}
\end{center}
\end{figure}

Figure \ref{fig:3d_sphere_ellipsoid} shows the mapping of a sphere onto an ellipsoid. We observe the same limitations for minimal distance mapping in 3D (Fig. \ref{fig:3d_minimal_distance}) that we have seen already in 2D (Fig.\,\ref{fig:circle_min}). Much as in 2D, the sphere is not mapped to all regions of the ellipsoid. Once again normal mapping (Fig.\,\ref{fig:3d_normal_mapping}) and diffusion mapping (Fig.\,\ref{fig:3d_diffusion_mapping}) perform well. 

\begin{figure}
\begin{center}
\includegraphics[trim=10cm 1cm 8.7cm 1cm, clip=true, width=0.5\textwidth, angle = 90]{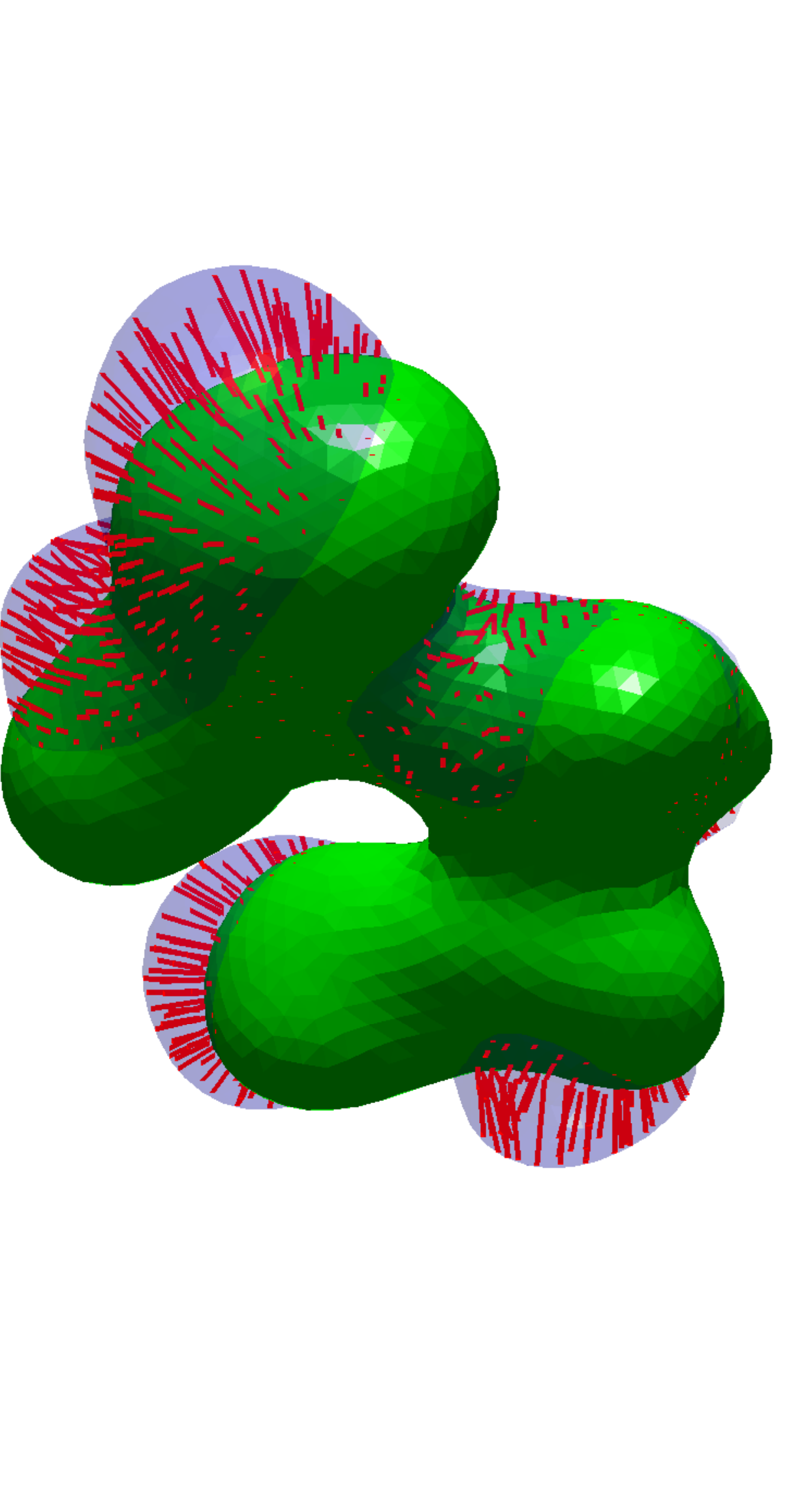} 
\caption{\textbf{Normal Mapping applied to 3D Lung Data.} The displacement field computed by using the normal mapping algorithm.}
\label{fig:lung}
\end{center}
\end{figure}

Figure\,\ref{fig:lung} shows the displacement field computed for the embryonic mouse lung epithelium at two developmental stages \cite{Blanc:2012ea}. Similarly as for the example of a sphere and an ellipsoid, normal mapping and diffusional mapping provide high quality displacement fields, as judged by eye.
  
\section{Conclusion}
Based on the evaluation of the mapping algorithms we propose a workflow (Algorithm\,\ref{alg:final} for the 2D case) 
for the determination of high quality displacement fields from images of subsequent developmental stages. From our overall observations and analysis we learn that it is generally desirable to apply normal mapping as a first attempt.   
When curves (in 2D) or surfaces (in 3D) intersect, it is advisable to divide the problem into sub-problems according to intersections and then apply the mapping algorithms separately on each segment.  {If we get a low quality displacement field (e\,.g. crossings occur), then the normal mapping must be rejected and reverse diffusion mapping can be used.} 
Reverse diffusion mapping is less error--prone when the two curves are far away. In the case of open curves, we can transform the starting and end points of both curves onto each other by translating, rotating and scaling one of the curves, then apply (scaled) normal or reverse diffusion mapping, depending on the crossings as for closed curve segments and finally  transform back the curve and the displacement field. \\ 

\begin{algorithm}[h]
\SetAlgoNoLine
\KwIn{Two closed curves $C_1$ and $C_2$}
\KwOut{Displacement field}
\eIf{$C_1$ and $C_2$ intersect}{
split into subproblems;}
{consider $C_1$ and $C_2$ as one subproblem}
\ForEach{subproblem}{
do normal mapping;

\If{crossings}{
do reverse diffusion mapping;
}
}
\Return{displacement field;}

\caption{ {Algorithm to determine 2D Displacement Fields}}
\label{alg:final}
\end{algorithm}

In this manuscript, we described and evaluated different algorithms that can be used to compute displacement fields that are required for the simulation of signalling models on growing domains. We have presented four basic algorithms, discussed their properties and presented an algorithm that takes their advantages and disadvantages into account. We must note that our final algorithm does not guarantee a perfect displacement field, nor that the best basic method is used. However, the algorithm works well for many cases and  minimises manual curation and modifications of the displacement field, as intended.

 \paragraph{Acknowledgements} We thank Edoardo Mazza, Gerald Kress and Simon Tanaka for discussions.

\bibliographystyle{ACM-Reference-Format-Journals}
\bibliography{Library_Papers}

\appendix
\section*{APPENDIX A} \label{app:transformation}

The details for the transformation algorithm used in section \ref{sec:extensions} are as follows: 
\begin{itemize}
\item compute the area for both $C_1$ and $C_2$, then scale $C_1$ around its centre of mass to the same area as $C_2$;
\item apply rigid transformation to $C_{1,scaled}$ so that the overlapping area   {of $C_2$ and $C_{1, scaled}$ is maximized.}
Finding an optimal translation and rotation is computationally very expensive and so alternatively we search for the solution using a greedy algorithm and do not go through all possible curve configurations but rather move the curve step by step to the next best configuration until no local improvement can be made (see Algorithm \ref{alg:transformation}). Therefore, we might not find the global optimal solution, but just a locally optimal solution, which, for our purpose, is good enough. We can then use any of the above mentioned algorithms (1) -- (4) to compute the mapping from $C_{1,t}$ to $C_2$. The starting points of these displacement field vectors are then transformed back onto $C_1$.
 \end{itemize} 
  
\begin{algorithm}[h]
\SetAlgoNoLine
\KwIn{Curve $C_1$ which is given by $N$ points,  {curve $C_2$}, step size $r$, rotation angles $\theta$}
\KwOut{Curve $C_{1,transformed}$ which consists of $N$ transformed points}
$a_1 =$ area\_of\_curve($C_1$), 
$a_2 =$ area\_of\_curve($C_2$)\;
$C_{1,scaled} = $ scale\_Curve($C_1$,$\frac{a_2}{a_1}$)\;
$A_{opt} = $ curve\_intersection\_area($C_{1,scaled}$,$C_2$)\;
$improvement = TRUE$\;
\Repeat{ {improvement is FALSE}}{
	$improvement = FALSE$\;
	\For{each translational direction $\phi$}{
		\For{$sign \in \{-1, 1\}$}{
			$ {A_{\theta,old}}$ = 0\;
			\For{each rotation angle $\theta$}{
			
				$C_{1,temp} = $ rotate\_curve($C_{1,scaled}$, $sign\cdot \theta$)\;
				$C_{1,temp} = $ translate\_curve($C_{1,temp}$,$r$,$sign \cdot \phi$)\;
				$A_{\theta,new} = $ curve\_intersection\_area($C_{1,temp}$,$C_2$)\;
				$\Delta Area = A_{\theta,new} -  A_{opt}$\;
				\If{$\Delta Area < 0 $}{
					$A_{opt} = A_{temp}$\;
					$improvement = TRUE$\;	
				}	

				\If{$Area_{\theta,new} - Area_{\theta,old} < 0$}{
					break\;
				}
				$Area_{\theta,old} = Area_{\theta,new}$\;

			}
		}
		
	}
}
\caption{Transformation of Curves}
\label{alg:transformation}
\end{algorithm}

\end{document}